\documentclass[%
 reprint,
 amsmath,amssymb,
 aps,
]{revtex4-2}

\usepackage{graphicx}
\usepackage{dcolumn}
\usepackage{bm}

\usepackage{acronym} 
\usepackage{url}
\usepackage{multirow}
\usepackage{amsmath}
\usepackage{amssymb}
\usepackage{amsfonts}
\usepackage{xcolor}
\usepackage{acronym}
\usepackage{ulem}
\usepackage{multirow}
\usepackage{booktabs}
\usepackage[export]{adjustbox}
\usepackage{algorithmic}
\usepackage{algorithm}
\usepackage{comment}

\usepackage{subfigure}

\graphicspath{%
  {results_plot_2026Feb/}{2026-02-01-PE-comparision-Nbran5-flow-9-dim-run2/}
}
\begin{document}

\newcommand{\xt}[1]{{\textcolor{cyan}{{\sf{[Xueting: #1]}}} }}
\newcommand{\nk}[1]{{\textcolor{purple}{{\sf{[Natalia: #1]}}} }}
\newcommand{\jg}[1]{{\textcolor{orange}{{\sf{{[Jon}: #1]}}} }}
\newcommand{\jgmod}[1]{{\textcolor{blue}{{#1}}}}
\newcommand{\yh}[1]{{\textcolor{brown}{{\sf{[YMHu: #1]}}} }}

\newcommand{\phenomd}{\texttt{IMRPhenomD}}
\newcommand{\red}{\textcolor{red}}
\newcommand{\Msun}{M_{\odot}}
\newcommand{\e}{\mathrm{e}}
\newcommand{\iu}{\mathrm{i}}
\newcommand{\tc}{t_{\mathrm{c}}}

\newcommand{\tabref}[1]{Tab. \ref{#1}}
\renewcommand{\eqref}[1]{Eq. (\ref{#1})}
\newcommand{\figref}[1]{Fig.~\ref{#1}}
\newcommand{\secref}[1]{Sec.~\ref{#1}}
\newcommand{\appref}[1]{Appendix \ref{#1}}


\newcommand{\smartpdf}[2][]{%
  \IfFileExists{#2}%
    {\immediate\write18{pdfcrop "#2" "\jobname-crop.pdf"}%
     \includegraphics[max width=\linewidth, max height=.9\textheight, keepaspectratio,#1]{\jobname-crop.pdf}}%
    {\includegraphics[max width=\linewidth, max height=.9\textheight, keepaspectratio,#1]{#2}}%
}

\newcommand{\TRC}{TianQin Research Center for Gravitational Physics, Sun Yat-sen University, 2 Daxue Rd., Zhuhai 519082, China}
\newcommand{\SPA}{School of Physics and Astronomy, Sun Yat-sen University, 2 Daxue Rd., Zhuhai 519082, China}
\newcommand{\SUPA}{School of Physics and Astronomy, University of Glasgow, Glasgow G12 8QQ, United Kingdom}
\newcommand{\AEI}{Max Planck Institute for Gravitational Physics (Albert Einstein Institute), D-14476 Potsdam, Germany}
\newcommand{\OCA}{Universit\'e C\^ote d’Azur, Observatoire de la C\^ote d’Azur, CNRS, Artemis, Bd de l’Observatoire, F-06304 Nice, France}

\preprint{APS/123-QED}

\title{Pre-localization of Massive Black Hole Binaries in the Millihertz Band}

\author{Xue-Ting Zhang}
\altaffiliation{\AEI}
\altaffiliation{\TRC} 
\altaffiliation{\SPA}

\email{xueting.zhang@aei.mpg.de}


 \author{Jonathan Gair}
 \altaffiliation{\AEI}

 \author{Chris Messenger}
\altaffiliation{\SUPA}

\author{Natalia Korsakova}
\altaffiliation{\OCA}

\author{Yi-Ming Hu}
\altaffiliation{TianQin Research Center for Gravitational Physics, Sun Yat-sen University, 2 Daxue Rd., Zhuhai 519082, China}
\altaffiliation[Also at ]{School of Physics and Astronomy, Sun Yat-sen University, 2 Daxue Rd., Zhuhai 519082, China}
\email{Email:huyiming@mail.sysu.edu.cn}

\author{Hong-Yu Chen}
\altaffiliation{TianQin Research Center for Gravitational Physics, Sun Yat-sen University, 2 Daxue Rd., Zhuhai 519082, China}
\altaffiliation[Also at ]{School of Physics and Astronomy, Sun Yat-sen University, 2 Daxue Rd., Zhuhai 519082, China}

\date{\today}

\begin{abstract}
The space-borne gravitational-wave (GW) detectors will open a new mass and redshift regime, allowing us to observe massive black hole binaries (MBHBs) throughout the Universe. 
A subset of these systems is expected to produce electromagnetic (EM) counterparts, offering a unique opportunity to follow the continuous evolution of massive black holes through joint GW and EM observations. 
Realizing this potential, however, requires low-latency, high-throughput data-analysis pipelines that can extract reliable source parameters and sky localizations from space-borne data streams fast enough to trigger EM follow-up.
In this work we develop a fast, normalising flow-based inference pipeline designed for early-warning analysis of MBHB signals in a TianQin-like configuration. 
Our method combines a learned embedding of the detector time series with a neural spline flow~(NSF) to perform amortized Bayesian inference, producing posterior samples for the main source parameters in roughly one minute per event. 
For a representative MBHB whose merger occurs $\sim 15$ minutes after the end of the analyzed GW observation, the pipeline achieves pre-merger sky localizations of order $\sim 20~\mathrm{deg}^2$, 
recovers the same number of sky modes as a reference parallel-tempered Markov chain Monte Carlo (PTMCMC) analysis, and yields parameter uncertainties of comparable scale, while still operating within a practically useful pre-merger warning window.
These results demonstrate that NSF-based inference can deliver accurate, near-real-time  
parameter estimation for space-borne MBHB GW signals, and that the resulting early-warning localizations are sufficiently precise to make rapid EM follow-up.


\acrodef{GW}[GW]{gravitational wave}
\acrodef{MBH}[MBH]{massive black hole}
\acrodef{sBH}{stellar-mass black hole}
\acrodef{MBHB}[MBHB]{massive black hole  binary}
\acrodef{sBBH}{stellar-mass binary black hole}
\acrodef{CO}[CO]{compact object}
\acrodef{DWD}[DWD]{double white dwarf}
\acrodef{BBH}[BBH]{binary black hole}
\acrodef{SOBH}[]{stellar origin black hole}
\acrodef{BNS}[BNS]{binary neutron star}
\acrodef{BH}[BH]{black hole}
\acrodef{NS}[NS]{neutron star}
\acrodef{WD}[WD]{white dwarf}
\acrodef{EMRI}[EMRI]{extreme mass ratio inspiral}
\acrodef{AGN}{active galactic nuclei}

\acrodef{SNR}[SNR]{signal-to-noise ratio}

\acrodef{TDI}{time delay interferometry}
\acrodef{PSD}{power spectral density}
\acrodef{MMA}{Multi-Messenger Astronomy}
\acrodef{mHz}{millihertz}

\acrodef{CNN}{convolutional neural network}
\acrodef{SBI}{Simulation-based inference}
\acrodef{LFI}{likelihood-free inference}
\acrodef{EM}{electromagnetic}
\acrodef{NSF}{Neural spline flow}
\acrodef{PTMCMC}{Parallel Tempered Markov chain Monte Carlo}
\acrodef{HPD}{highest-posterior-density}
\acrodef{KDE}{kernel density estimation}
\acrodef{KNN}{K-nearest neighbor}

\end{abstract}

\maketitle


\section{\label{sec:level1} Introduction}

The first detection of \acp{GW} from a binary black hole  merger in 2015 opened the era of GW astronomy \citep{LIGOScientific:2016aoc}.
Future space-borne GW detectors, such as TianQin~\cite{luo2016tianqin,Luo:2020bls} and LISA~\citep{Amaro-Seoane:2012aqc,LISA:2022yao}, will extend this reach to much lower frequencies and higher masses, making it possible to observe a \ac{MBHB} with component masses in the range of $10^{4}$--$10^{8}\,M_{\odot}$ out to cosmological distances~\citep{Wang:2019ryf,Bogdanovic:2021aav}. 
A subset of these systems is expected to reside in gas-rich galactic nuclei and to power luminous \ac{AGN}, relativistic jets, and other forms of \ac{EM} emission~\cite{Xin:2024fci,Krauth:2025tur}. 
For such sources, joint GW and EM observations offer a unique opportunity to follow the assembly and coalescence of massive black holes across cosmic time~\citep{Bogdanovic:2021aav,Mangiagli:2022niy,Xin:2024fci}.

The multi-messenger potential of MBHBs is particularly compelling, but also fragile. 
Population studies combining GW and EM selection effects indicate that only a small fraction of MBHB mergers detectable in the \ac{mHz} band will produce EM counterparts bright enough to be identified with current or near-future facilities, with typical joint detection rates of order a few events over a mission lifetime~\citep{Mangiagli:2022niy,Xin:2024fci}. 
This rarity implies that each multi-messenger MBHB event carries high scientific value, and motivates dedicated efforts to optimise the GW analysis for EM follow-up rather than treating multi-messenger observations as an incidental by-product.

The EM phenomenology expected from gas-embedded MBHBs spans a wide range of timescales \cite{Lai:2022ylu,DeLaurentiis:2024rbc}. 
Hydrodynamical and general-relativistic magnetohydrodynamics simulations of circumbinary accretion flows predict that systems in gas-rich environments generically exhibit strong time variability, including episodes of enhanced accretion, shock heating, and abrupt changes in luminosity in the final days to hours before merger~\citep{DeLaurentiis:2024rbc,Lai:2022ylu,Krauth:2023nyq}. 
In one regime, eccentric streams and mini-discs can produce repeated flares or strongly modulated emission on orbital timescales, with individual events lasting from tens of minutes to hours but recurring over many cycles \cite{Lai:2022ylu,DeLaurentiis:2024rbc}. 
For such signals, the dominant requirement on the GW side is accurate pre-merger localization in sky position and luminosity distance, together with a rough estimate of the merger time, so that EM facilities can concentrate high-cadence monitoring on a small number of candidate AGN in the last few orbits before coalescence.
In a more demanding regime, several models predict short-lived, phase-locked flares whose occurrence is tied to a narrow range of orbital phases, such as self-lensing flares when one black hole lenses the mini-disc around its companion or other sharp spikes associated with late-stage accretion and jet activity ~\citep{Davelaar:2021eoi,Krauth:2023svh,Ingram:2021gar,DeLaurentiis:2024rbc}. 
These flares are scientifically rich because their morphology and timing relative to the GW chirp encode detailed information about the binary geometry, inclination, and disc structure, and can break degeneracies that are difficult to address with GW data alone~\citep{Krauth:2023svh,Dong-Paez:2023qlf}. 
However, their characteristic durations range from minutes to hours, and they often occur only once or a few times near coalescence, which makes them easy to miss unless GW-based early-warning information is available with sufficiently low latency and high fidelity.

From the GW perspective, MBHB signals in the \ac{mHz} band are both promising and challenging. 
On the one hand, their low frequencies ($\sim 10^{-4}$--$10^{-1}\,\mathrm{Hz}$) and long durations in band allow space-based GW detectors to accumulate very high SNRs, often 
$\gtrsim 10^{2}$--$10^{3}$, with the bulk of the SNR acquired in the final day before coalescence~\citep{Wang:2019ryf,Feng:2019wgq}. 
In principle, this enables extremely precise parameter estimation, including tight constraints on the sky location, luminosity distance, and coalescence time~\citep{Bogdanovic:2021aav}. 
On the other hand, extracting this information from continuous space-borne data streams is computationally expensive: the signals live in a high-dimensional parameter space, coexist with other sources and instrumental backgrounds, and require repeated waveform evaluations over long observation windows. 
Standard Bayesian inference schemes based on \ac{PTMCMC} or nested sampling can therefore take hours to days per event ~\citep{Hoy:2023ndx}, which is incompatible with genuine early-warning operation for short-lived EM flares.

A variety of strategies have been proposed to accelerate MBHB parameter estimation. 
One broad class focuses on speeding up likelihood evaluations, for example through heterodyned likelihoods~\citep{Cornish:2021lje,Chen:2023qga,Katz:2024oqg} and related techniques. 
A second class uses likelihood-free inference based on generative models, such as normalizing flows or other conditional density estimators, to amortize the cost of parameter estimation over many events~\citep{Vilchez:2024qnw,Du:2023plr,Spadaro:2026evb}. 
A third class improves sampling efficiency by constructing better proposal distributions or narrowing the prior to a reduced region of parameter space, often in hierarchical multi-stage schemes~\citep{Hoy:2023ndx,Weaving:2023fji}. 
On the hardware side, GPU-based parallelization has also been explored to reduce wall-clock time~\citep{Katz:2024oqg}. 
These approaches have achieved substantial gains in specific regimes; however, none of the previous studies has focussed on the most difficult aspects relevant to short-lived EM flares,  namely: 
(i) delivering reliable, fully Bayesian posteriors for sky location and coalescence time on minute timescales after an early-warning trigger, while
(ii) preserving multimodal structure in the posterior, and 
(iii) doing so in a data analysis framework appropriate to realistic space-borne detectors.

In this work we focus on precisely this regime. 
Our scientific goal is to enable GW early warning for MBHB systems that may produce short-lived, non-periodic, high-energy EM flares near merger. 
We ask whether it is possible to strike a practical balance between sky-localization accuracy and latency, such that EM facilities have a realistic chance of capturing these fast transients. 
To this end, we develop a normalizing-flow--based inference pipeline tailored to a TianQin-like configuration.
The pipeline combines a learned embedding of the detector time series with a \ac{NSF}, and is trained to perform amortized Bayesian inference for a reduced set of six parameters that are most directly relevant for EM follow-up: 
the chirp mass $\mathcal{M}_c$, symmetric mass ratio $\eta$, ecliptic longitude $\lambda$, ecliptic latitude $\beta$, coalescence time $t_c$, and luminosity distance $D_L$.The remaining source parameters are randomized over their priors during training but are not inferred explicitly, so that the model focuses its capacity on the quantities most critical for rapid localization and triggering.

Although the present implementation is trained for a TianQin-like setting, the underlying strategy is more general. The embedding network provides a detector-dependent summary of the time series, while the flow models the corresponding conditional posterior. Detector specificity therefore enters mainly through the simulated data and training distribution, including the instrument response, noise properties, and observing configuration, rather than through the inference architecture itself. In this sense, the method is best viewed as a transferable inference framework that can be adapted to other detectors through retraining under the appropriate instrument model, including space-based observatories such as 
LISA~\cite{LISA:2024hlh} and Taiji~\cite{Luo:2021qji}. Finally, 
the normalizing flow is validated
against a \ac{PTMCMC} baseline, which demonstrates that the normalizing flow correctly preserves multimodal posterior structure and avoids bias in the early-warning regime.
The resulting pipeline demonstrates that rapid and calibrated posterior estimation is feasible on timescales relevant for multimessenger follow-up, providing a practical building block for future low-latency GW--EM searches.

This paper is organized as follows. In 
Sec.~\ref{subsec:mbhb-sig-gen}, we describe the simulation setup and characteristics of the MBHB signals considered. Sec.~\ref{subsec:mbhb-relative-binning} presents the \ac{PTMCMC} sampler accelerated with heterodyned likelihoods, which serves as a baseline for comparison. Sec.~\ref{subsec:mbhb-nsf} introduces our likelihood-free inference approach based on \ac{NSF} and the associated embedding network. Finally, in Sec.~\ref{subsec:mbhb-results}, we compare the \ac{NSF} and \ac{PTMCMC} posteriors, quantifying the accuracy, latency, and robustness, and assess the implications for EM follow-up. 
Finally, Sec.~\ref{sec:conclusions} summarizes our findings and outlines directions for future work towards a practical low-latency \ac{GW} early-warning pipeline for multimessenger follow-up.. 

\section{Mock data}\label{subsec:mbhb-sig-gen}
In this section we describe how the datasets used for training and validation were constructed.




\subsection{Massive black hole binary \ac{GW} signal}\label{subsec:tianqin-channels}




A space-based interferometer, like TianQin, comprises three spacecraft forming a quasi-equilateral triangle; laser links between each pair measure differential optical path-length fluctuations induced by passing GWs.
The \ac{TDI} cancels laser frequency noise and yields three approximately uncorrelated data streams with diagonal noise covariance: the canonical TDI-A, TDI-E, and TDI-T channels~\citep{Tinto:2004yw,2001CQGTinto,1999Tinto,Wang:2022nea}.

When a MBHB transits the detector band, the data stream contains both instrument noise and the GW response modulated by the detector motion and geometry.
Early studies often modeled this response using only the dominant $(\ell,m)=(2,2)$ mode (e.g., \texttt{IMRPhenomD}~\citep{Ajith:2009bn}). 
In the early-warning regime, as the system approaches merger, the amplitude and instantaneous frequency evolve rapidly, which tightens the accuracy requirements on waveform modeling.
Higher-order modes are known to improve parameter estimation, in particular sky localization, in LISA-band analyses~\cite{Marsat:2020rtl}.
Accordingly, we employ an aligned-spin phenomenological waveform  \texttt{IMRPhenomHM} \cite{London:2017bcn} which includes the dominant $(2,2)$ mode and the subdominant higher-order modes $(2,1)$, $(3,3)$, $(3,2)$, $(4,4)$ and $(4,3)$.
At earlier times, far from merger, the higher-order modes become progressively weaker and may carry too little \ac{SNR} to be clearly resolved, so the signal is expected to be dominated by the $(2,2)$ mode \citep{Marsat:2020rtl}.
This is consistent with previous studies of the early inspiral phase, which often focused only on the dominant mode \citep{Chen:2024egt,Chen:2023qga,Marsat:2020rtl}.

Let $d^{C}(t)$ denote the TDI channel, i.e., $C\in\{\rm{A,E,T}\}$ and $n^{C}(t)$ its noise.
The measured data can be written as a superposition of the mode responses plus noise,
\begin{equation}
d^{C}(t) \;=\;  s^{C}(t) \;+\; n^{C}(t).
\end{equation}
Each TDI stream is a linear combination of delayed link measurements $s^{C}(t) =f^C( y_{sr})$, which is represented as a combination of single-link observables, $ y_{sr}$; in turn, each single-link observable is the detector response to the incident metric perturbation along the link.
Schematically, for link $y_{sr}$ (from spacecraft $s$ to $r$) one may write the time-domain link response for a given $(\ell,m)$ mode as
\begin{equation}\label{eq:single-link}
y_{sr}(t)
= \frac{1}{2}\,
\frac{\hat{n}_{sr}\otimes \hat{ n}_{sr}}{1-\hat{ k}\!\cdot\!\hat{n}_{sr}} :
\Big[\,H_{\ell m}\!\big(t - \frac{L}{c} - \frac{ \hat{k} \!\cdot\! \hat{p}_s}{c} \big)
- H_{\ell m}\!\big(t - \frac{\hat{  k}\!\cdot \! \hat{p}_r}{c}\big)\,\Big],
\end{equation}
where $c$ denotes the speed of light, $\hat{ k}$ is the GW propagation unit vector, 
$\hat{ n}_{sr}$ the unit vector along the link from $s$ to $r$, 
$\hat{p}_{s,r}$ the spacecraft positions, and ${H}_{\ell m}$ the metric perturbation contribution of the $(\ell,m)$ mode (in TT gauge). Additionally, the length of the link from $s$ to $r$ is denoted $L_{sr}$. 
The TDI channels are synthesized by appropriate delay-and-combine operations on any signal mode $y_{sr}$. 
We use $y_{sr, \ T_{sr}}$ to denote the single link measurement from  $s$ to  $r$, delayed by a time $T_{sr}$ corresponding to the arm length $L_{sr}$. 
The first generation TDI observables are TDI-X, TDI-Y, and TDI-Z with unequal arms. 
These are:
\begin{align*}
    \rm{X}(t) & =\big[ y_{31} + y_{13,\ T_{31}} + y_{21,\ T_{13}+T_{31}} + y_{12,\ T_{12} + T_{13} + T_{31}}  \big] \\
    &  - \big[y_{21} + y_{12,\ T_{21}} + y_{31,\ T_{12}+T_{21}}  +  y_{13,\ T_{31}+T_{12}+T_{21}}   \big],\\
    \rm{Y}(t) & = \big[y_{12} + y_{21,\ T_{12}} + y_{32,\ T_{21} + T_{12}}  + y_{23,\ T_{31}+T_{21}+T_{12}}  \big] \\
    &- \big[ y_{32} + y_{23, \ T_{32}} + y_{12, \ T_{22}+T_{32}} +y_{21, \ T_{12}+T_{23}+T_{32}}  \big] ,\\
    \rm{Z}(t) & = \big[y_{23} + y_{32,\ T_{23}} + y_{13,\ T_{32}+T_{23}}  + y_{31,\ T_{13}+T_{32}+T_{23}}  \big] \\
    &- \big[ y_{13} + y_{31, \ T_{13}} + y_{23, \ T_{31}+T_{13}} +y_{32, \ T_{32}+T_{31}+T_{13}}  \big]. 
\end{align*}
Since we assume they are the equal arms, the TDI observables can be simplified. For example, 
\begin{equation}
\begin{split}
        \rm{X}(t)  = & \ y_{31} + y_{13,\ L/c} + y_{21,\ 2L/c} + y_{12,\ 3L/c}  \\
    & - \big[ y_{21} + y_{12,\ L/c}  + y_{31,\ 2L/c} + y_{13, \ 3L/c} \big]
\end{split}
\end{equation}
The optimal TDI observables are constructed as linear combinations of the TDI-X, TDI-Y, and TDI-Z channels. 
\begin{align*}
    \rm{A}(t) & =\frac{1}{\sqrt{2}} (\rm{Z}-\rm{X}), \\
    \rm{E}(t) & = \frac{1}{\sqrt{6}} (\rm{X}-2\rm{Y}+\rm{Z}) , \\
    \rm{T}(t) & = \frac{1}{3}\big(\rm{X}+\rm{Y}+ \rm{Z} \big).
\end{align*}

Under the stationary, quasi-monochromatic (locally stationary) approximation, the response admits a frequency-domain representation in terms of transfer functions that depend on the detector configuration and the time–frequency evolution of the source.
Introducing the Fourier-domain GW mode amplitudes $\tilde h_{\ell m}(f)$ and the corresponding TDI transfer functions $\,\mathcal{T}^{C, \ell m}(f;\Omega)\,$  (with $\Omega$ collecting sky position, polarization, inclination, etc, which also means $\,\mathcal{T}$ is carrying information on the detector modulation and the time-delays of the laser links.), one obtains
\begin{equation}\label{eq:fd-model}
\tilde{a},\tilde{e},\tilde{t}  \;=\; \sum_{\ell m}   \mathcal{T}^{C,\ell m}(f;\Omega)\ \tilde h_{\ell m}(f). 
\end{equation}
Equivalently, one may express the delay associated with a given harmonic as
\begin{equation}
t_{\ell m}(f) \;=\; t_{\rm ref} - \frac{1}{2\pi}\,\frac{{\rm d}\Psi_{\ell m}(f)}{{\rm d}f},
\end{equation}
where $\Psi_{\ell m}(f)$ is the Fourier-domain phase of $\tilde h_{\ell m}(f)$ and $t_{\rm ref}$ is an arbitrary reference time.

Although the mode-decomposition form above \eqref{eq:fd-model} is generic to space-based TDI observables, the detector-specific information is encoded in $\mathcal{T}^{C,\ell m}(f;\Omega)$. In the present work this operator is evaluated for the TianQin configuration, rather than the standard LISA one. Quantitatively, TianQin is a geocentric constellation with orbital radius of order $10^5\,\mathrm{km}$ and a single-satellite orbital period of about $3.64\,\mathrm{days}$, and it typically operates with a ``3 months on + 3 months off'' observing cadence. 
By contrast, LISA is heliocentric and exhibits modulation on the characteristic annual timescale. This distinction has direct consequences for the response: LISA benefits from strong annual amplitude modulation below $\sim 1\,\mathrm{mHz}$, whereas TianQin, whose detector plane points toward a fixed reference source direction, has much weaker amplitude modulation in that regime. At higher frequencies, however, Doppler modulation becomes increasingly important, and TianQin can surpass LISA in sky localization above roughly $30\,\mathrm{mHz}$~\cite{Zhang:2020hyx}.
The details of $\mathcal{T}^{C,\ell m}(f;\Omega)$ can be found in the Python package \cite{Li:2023szq,gwspace}. 


If no GW is present, each TDI channel contains only instrument noise characterized by the one-sided \ac{PSD}.
For TianQin, adopting the standard equal-arm expressions, one has~\cite{Li:2023szq}
\begin{align}\label{eq:psd-AET}
S_{\rm{A,E}}(f)  \;=\;  &
8\,\sin^{2} f_c
[
(\cos f_c+2)\,S_{\rm{oms}}(f)    \\
 & + 2\big( \cos(2f_c)+    
 2\cos f_c+3 \big)\ S_{\rm{acc}}(f)
], \nonumber \\
S_{\rm{T}}(f)
 = & \ 32\,\sin^{2} f_c \,\sin^{2}\!\frac{f_c}{2}\,
\left[\,4\sin^{2}\! \frac{f_c}{2}\, S_{\rm{acc}}(f) + S_{\rm{oms}}(f) \right].
\end{align}
Here $S_{\rm{acc}}$ denotes the residual acceleration noise and $S_{\rm{oms}}$ the position (optical path) readout noise \citep{luo2016tianqin,Luo:2020bls}]. 

\subsection{Prior choices motivated by early-warning considerations}\label{subsec:early-warning}

%


The prior ranges adopted in this work are not intended to represent the full astrophysical MBHB population, but rather to isolate the part of parameter space most relevant for a practical joint GW--EM search. In this context, the systems of greatest interest are MBHBs embedded in gas-rich galactic nuclei, for which a range of possible \ac{EM} activity has been discussed~\cite{Xin:2024fci,Krauth:2025tur}, including short-lived flare-like events in the optical, UV, and X-ray bands \cite{Bogdanovic:2021aav}. Such signals are particularly challenging because their detectability depends not only on source brightness, but also on whether the GW localization can be updated rapidly enough to guide timely follow-up.


Detecting such flares in coincidence with MBHB mergers is only feasible if a real-time or low-latency gravitational-wave pipeline can provide accurate, rapidly updated posteriors for the coalescence time and sky location, so that \ac{EM} facilities can schedule timely, deep observations during the narrow window when these transients are most likely to occur. 
Otherwise, short-lived \ac{EM} counterparts to MBHB mergers will remain largely undetected, and the opportunity to complete the census of massive black holes through joint \ac{GW} and \ac{EM} observations will be severely limited.

This is the main reason for focusing on the late inspiral portion of the GW signal. The point is not to assume a one-to-one correspondence between late-time GW emission and any specific flare model, but to recognize that the main difficulty of \ac{EM} follow-up is often practical: identifying viable candidates within a limited response time and over a localization region that may still be large. In this regime, reducing the GW sky area can be more valuable than modest gains in warning lead time. Late-time GW data are particularly informative because the sky-localization information accumulates highly nonlinearly near coalescence. For TianQin MBHB studies, previous work has shown that the final hour before merger can contribute a dominant fraction of the total \ac{SNR}, reaching up to $\sim 90\%$ in representative cases \cite{Feng:2019wgq}. Since localization improves strongly with increasing \ac{SNR}, incorporating more of this late-time signal can dramatically reduce the sky area that must be searched electromagnetically, even though the available warning time becomes correspondingly shorter. In this sense, late-time analysis is valuable not because it guarantees an \ac{EM} counterpart, but because it maximizes the practical usefulness of GW information for rapid counterpart searches.


This strategy is most useful for sources that also lie in the part of parameter space to which TianQin is most sensitive. Previous studies have shown that TianQin can detect coalescing MBHBs with detector-frame masses around $10^{5}$--$10^{7}\,M_\odot$ with high SNR, while systems in the lower part of this range are especially favorable for pre-merger localization and early warning \cite{Feng:2019wgq,Wang:2019ryf}. At the same time, \ac{EM} follow-up becomes increasingly difficult at high redshift, so the multimessenger-relevant population is naturally biased toward moderate-redshift systems that are both detectable in the millihertz band and not too distant for realistic counterpart searches. These considerations directly motivate the reduced prior volume adopted in Table~\ref{tab:mbhb_priors}. 
We restrict the detector-frame chirp mass to $\mathcal{M}_c \in [2\times10^{5},10^{6}]\,M_\odot$, corresponding to the part of the TianQin-sensitive MBHB population in which strong late-time signal accumulation and useful pre-merger localization are both expected. 
We take the redshift to lie in $z\in[0.1,2]$, sampled uniformly in comoving volume, so as to emphasize the regime where GW detectability and the prospects for \ac{EM} identification still overlap~\cite{Mangiagli:2022niy}; the luminosity distance $D_L$ is then derived from $z$ and the assumed cosmology. Sky position and orientation are assigned standard isotropic priors, with $\lambda\in[0,2\pi)$, $\sin\beta\in[-1,1]$, $\cos\iota\in[-1,1]$, and $\psi\in[0,\pi)$, while the coalescence phase is taken as $\phi_c\in[0,2\pi)$. Finally, for a 5-day observation segment ending at $t_{\rm end}$ with a sample rate of 0.01 Hz, we impose
$
t_c - t_{\rm end} \sim \mathcal{U}(600,1200)\ {\rm s},
$
so that all training examples lie in a genuinely near-merger, low-latency regime while still leaving a short but nonzero interval for follow-up scheduling.

\begin{table*}[t]
\centering
\caption{
Priors adopted for the multimessenger-motivated MBHB population considered in this work. The observation time is 5 days. Here $t_{\rm end}$ denotes the end time of the observation segment, and the coalescence time is placed 10--20 minutes after $t_{\rm end}$.}
\label{tab:mbhb_priors}
\begin{tabular}{lll}
\hline
\textbf{Parameter} & \textbf{Prior range} & \textbf{Physical interpretation} \\
\hline
$\log_{10} ( \mathcal{M}_c [M_\odot])$ 
 &  Uniform in $[\log_{10}  (2\times 10^{5})$ , 6 ]
 & Detector-frame chirp  mass \\[2pt]

$\eta [-]$ 
 & Uniform in the $[0.05, 0.25]$ 
 & Symmetric mass ratio \\[2pt]

$z [-]$ 
 & Uniform in the co-moving volume [$0.1$, $2$]
 & Cosmological redshift   \\[2pt]

$D_L [\rm Gpc]$ 
 & Derived from $z$ and cosmology  
 & Luminosity distance\\[2pt]

$\lambda [\rm rad]$ 
 &  Uniform in $[0,2\pi)$
 &  Ecliptic latitude \\[2pt]

$\sin (\beta [\rm rad])$ 
 &  Uniform in $[-1,1]$ 
 & Sin ecliptic latitude \\[2pt]
 
$\cos (\iota [\rm rad])$ 
 & Uniform in $[-1,1]$ 
 & Binary inclination \\[2pt]

$\psi [\rm rad]$ 
 & Uniform in $[0 , \pi)$ 
 & Polarization angle \\[2pt]

 
$\Delta t_c \,[\mathrm{s}] =t_c - t_{\rm end}$
 & Uniform in $[600,\,1200]$
 & Time from observation end to coalescence \\[2pt]

$\phi_c [\rm rad]$ 
 & Uniform in $[0,2\pi)$  
 & Orbital phase at coalescence\\[2pt]


\hline
\end{tabular}
\end{table*}

\section{Likelihood-based parameter inference}\label{subsec:mbhb-relative-binning}





\subsection{Bayes' theorem  and  Parallel Tempered Markov chain Monte Carlo}\label{subsec:bayes-sampling}


For \ac{GW} parameter estimation, the posterior distribution based on Bayes' theorem is
\begin{equation}\label{eq:bayes}
  p(\boldsymbol{\theta}\mid d)
  \;\propto\;
 \frac{ \mathcal{L}(d\mid \boldsymbol{\theta})\,\pi(\boldsymbol{\theta})}{\int \mathcal{L}(d\mid \boldsymbol{\theta})\,\pi(\boldsymbol{\theta})\,\mathrm{d}\boldsymbol{\theta}},
\end{equation}
where $\pi(\boldsymbol{\theta})$ is the prior and $\mathcal{L}(d\mid \boldsymbol{\theta})$ is the likelihood. The proportionality constant is the inverse of the Bayesian evidence, also known as the marginal likelihood.
Assuming stationary, Gaussian noise with one-sided \ac{PSD} $S_n(f)$, the log-likelihood for a template
$h(\boldsymbol{\theta})$ and data $d$ can be written in the standard noise-weighted inner-product form~\citep{Finn:1992wt}
\begin{align}\label{eq:likelihood}
  \log \mathcal{L}(d\mid \boldsymbol{\theta})
  &\;=\; 
  -\tfrac{1}{2}\,\langle d-h(\boldsymbol{\theta}) \mid d-h(\boldsymbol{\theta}) \rangle + \mathrm{constant} \nonumber \\
  & \;=\; 
  \langle d \mid h(\boldsymbol{\theta}) \rangle - \tfrac{1}{2}\,\langle h(\boldsymbol{\theta}) \mid h(\boldsymbol{\theta}) \rangle  + \mathrm{constant},
\end{align}
where
\begin{equation}
  \langle a \mid b \rangle \equiv
  4\,\mathfrak{Re}\!\int_{f_{\min}}^{f_{\max}}
  \frac{\tilde a(f)\,\tilde b^{\,*}(f)}{S_n(f)}\,\mathrm{d}f.
\end{equation}
Strictly speaking, the additive constant is part of the fully normalized likelihood. However, in parameter estimation, one is often interested only in likelihood ratios, or posterior shapes up to normalization. Since this constant is independent of the source parameters, it does not affect the inference and is therefore usually omitted in practice.

When the likelihood is intractable to optimize analytically or the posterior exhibits a complicated geometry, stochastic
sampling is employed to approximate $p(\boldsymbol{\theta}\mid d)$. Two widely used techniques are
(i) Markov Chain Monte Carlo (MCMC); and (ii) nested sampling (which additionally yields the evidence)~\cite{2004AIPC..735..395S}.
MBHBs observed over short time windows tend to yield multi-modal posteriors ~\cite{Marsat:2020rtl}:
 fewer observed GW cycles and diminished spatial modulation sensitivity in the detector response reduce disambiguation power, increasing the number
and separation of posterior modes \cite{Marsat:2020rtl} 
To robustly traverse these separated modes we use \ac{PTMCMC}. 

\ac{PTMCMC} runs $K$ chains at different inverse temperatures $\beta_k \in (0,1]$,
sampling the tempered posteriors
\begin{equation}
  \pi_{\beta_k}(\boldsymbol{\theta})
  \;\propto\;
  \mathcal{L}(d\mid \boldsymbol{\theta})^{\beta_k}\,\pi(\boldsymbol{\theta}).
\end{equation}
The cold chain ($\beta=1$) targets the true posterior, whereas hot chains ($\beta \ll 1$) flatten the log-likelihood landscape and can therefore cross barriers between separated modes more easily \cite{Earl2005PCCP,Karnesis:2023ras}.
Information is exchanged by proposing replica swaps between adjacent temperatures.
If chains $i$ and $j$ (with $\beta_i>\beta_j$) currently sit at $\boldsymbol{\theta}_i$ and $\boldsymbol{\theta}_j$,
a state swap between $\boldsymbol{\theta}_i$ and $\boldsymbol{\theta}_j$ is accepted with Metropolis probability~\cite{Karnesis:2023ras} 
\begin{align}
  \alpha_{\rm swap} 
  &\;=\;     \nonumber
  \min\!\Big\{1,\,
  \frac{\mathcal{L}(d\mid \boldsymbol{\theta}_i)^{\beta_j}\,
        \mathcal{L}(d\mid \boldsymbol{\theta}_j)^{\beta_i}}  
       {\mathcal{L}(d\mid \boldsymbol{\theta}_i)^{\beta_i}\,
        \mathcal{L}(d\mid \boldsymbol{\theta}_j)^{\beta_j}}\Big\} \\ \nonumber
  &\;=\; 
  \min\!\Big\{1,\, \\
 & \exp\!\big[(\beta_i-\beta_j) \big(\ln \mathcal{L}(d\mid \boldsymbol{\theta_j}) 
   -\ln \mathcal{L}(d\mid \boldsymbol{\theta_i})\big)\big]\Big\}.
\end{align}
Periodic swaps allow hot chains to seed new modes that percolate down
to the cold chain, thereby improving global mixing while preserving detailed balance of the joint tempered ensemble.
The practical efficiency of \ac{PTMCMC} is governed by three related quantities: the adjacent-chain swap acceptance, which controls diffusion along the temperature ladder; the round-trip time, defined as the typical time required for a replica to move from the cold chain to the hottest chain and then return; and the autocorrelation time of the cold chain, which determines the effective number of independent posterior samples \cite{Katzgraber:2006,2016MNRAS.455.1919V}.

As the posterior becomes more multimodal or the barriers between modes grow, the round-trip time 
between hot and cold
chains and the overall convergence time typically increases.
Consequently, relying on \ac{PTMCMC} alone for low-latency early warning is impractical. 
However, convergence can be improved by including
optimized temperature ladders (targeting swap acceptance $\sim 0.2$–$0.5$), more effective proposal distributions (e.g., differential-evolution moves, Fisher-preconditioned steps
, sky-position proposals tailored to known directional degeneracies ~\cite{Marsat:2020rtl}), 
parallelization, and reduced-cost likelihood evaluations (e.g., heterodyned likelihoods ~\citep{Zackay:2018qdy,Cornish:2021lje}). 
These methods can substantially accelerate exploration of multimodal posteriors, although they do not by themselves remove the fundamental latency limitations of \ac{PTMCMC} in the early-warning regime.

\subsection{Heterodyned likelihood and validation}



To accelerate posterior sampling, an effective strategy is to construct a fast {approximate likelihood} that is faithful over the region of interest, thereby reducing the overall cost of Bayesian inference. 
The heterodyned, or relative-binning, likelihood achieves this by reducing the number of frequency points at which the likelihood must be evaluated explicitly \cite{Zackay:2018qdy,Cornish:2021lje}.
Its key idea is to preselect a reference template $\tilde{h}_0$ during an early screening phase and to evaluate candidate waveforms through their smooth frequency-dependent ratio to $\tilde{h}_0$.  When the signal of interest $\tilde{h}$ is sufficiently close to $\tilde{h}_0$, inner products that define the likelihood can then be approximated accurately in terms of the template ratio using a small number of frequency bins rather than the full frequency grid.

 We work with the three TDI channels $C\in\{A,E,T\}$. For notational simplicity, we suppress the channel label in the equations below.
For a trial parameter point $\theta$,
we define the complex residual ratio between the target and reference templatesin channel $C$ as
\begin{equation}\label{eq:r_def}
    \tilde r(f,\theta) 
  \equiv
  \tilde r_C(f,\theta) 
  \equiv
  \frac{\tilde h_C(f,\theta)}{\tilde h_{0,C}(f,\theta_0)}\,   ,
\end{equation}
where $\tilde h_{0,C}(f,\theta_0)$ is the reference waveform in channel $C$ evaluated at the fixed reference parameters $\theta_0$.
We partition the frequency axis into coarse bands $\mathcal B_j$ with centres $\bar{f}_b =\frac{1}{2}(\mathcal{B}_j+\mathcal{B}_{j+1})$.
The bands are chosen such that, within each band $b$, $\tilde r(f, \theta)$ is well approximated by a low-order smooth expansion; to linear order one writes
\begin{equation}\label{eq:r_tilde_f}
  \tilde r(f,\theta) \;=\; \tilde r_{0}^b(\theta) \;+\; \tilde r_{1}^b(\theta)\,\big(f - \bar{f}_b\big) \;+\; \mathcal{O}\!\big((f-\bar{f}_b))^2\big)\,,
  \qquad f\in b,
\end{equation}
where the complex coefficients $\bigl(\tilde r_{0}(b, \theta),\tilde r_{1}(b,\theta)\bigr)$ are band-wise constants determined by the pair $(h(f,\theta),h_0({f,\theta_0}))$. 

In the log-likelihood, Eq.~(\ref{eq:likelihood}), for data $d$ and template $h(\theta)$, the two components that need to be calculated are $\langle d| h(\theta) \rangle$ and 
$\langle h (\theta)| h (\theta) \rangle $.
Writing $\tilde h(f , \theta)=\tilde r(f , \theta)\,\tilde h_0(f,\theta_0)$ and inserting ~\eqref{eq:r_tilde_f} yields a fast approximation for the cross term
\begin{equation}\label{eq:Z_def}
\begin{split}
  \langle d \mid h(\theta) \rangle
  & \;\approx\;
  Z[\tilde d(f),\tilde h(f,\theta)] \\ &
  \;\equiv\;
  4\, \mathfrak{Re}  \!\sum_{i}
  \frac{\tilde d(f_i)\,\tilde h_0^{\,*}(f_i,\theta_0)}{S_n(f_i)}\;\tilde r(f_i, \theta)\,\Delta f,
  \end{split}
\end{equation}
where $\{f_i\}$ denote the dense frequency grid.
Using the banded linear expansion, ~\eqref{eq:Z_def} reduces to
\begin{equation}\label{eq:Z_banded}
  Z[\tilde d(f),\tilde h(f,\theta)]
  \;=\;
  4\,\mathfrak{Re}\!\sum_{ b\in \mathcal{B}}
  \Big[\,A_0(b)\,\tilde r_{0}^{\,*}(b,\theta) \;+\; A_1(b)\,\tilde r_{1}^{\,*}(b,\theta)\,\Big],
\end{equation}
where the (precomputable) band coefficients depend only on $(d,h_0)$:
\begin{align}
  A_0(b) \;&\equiv\; \sum_{f \in \mathcal{B}} \frac{\tilde d(f)\,\tilde h_0^{\,*}(f)}{S_n(f)}\,\Delta f, \\
  A_1(b) \;&\equiv\; \sum_{f\in \mathcal{B}} \frac{\tilde d(f)\,\tilde h_0^{\,*}(f)}{S_n(f)}\,(f-\bar{f}_b)\,\Delta f.
\end{align}
An analogous reduction applies to the auto term,
\begin{align}\label{eq:hh_banded}
  \langle h(\theta) \mid h(\theta) \rangle
  \;\approx\;
   Z[\tilde h,\tilde h] & =
  4\,\mathfrak{Re}\!\sum_{b \in \mathcal{B}}
  \Big[\,B_0(b)\,|\tilde r_{0}(b,\theta)|^2  \nonumber  \\  &
  \;+\; 2\,B_1(b)\,\mathfrak{Re} \big(\tilde r_{0}(b,\theta)\,\tilde r_{1}^{\,*}(b,\theta)\big)\Big],
\end{align}
with band coefficients that depend only on $h_0$:
\begin{align}
  B_0(b) \;&\equiv\; \sum_{f\in \mathcal{B}} \frac{|\tilde h_0(f)|^2}{S_n(f)}\,\Delta f, \\
  B_1(b) \;&\equiv\; \sum_{ f \in \mathcal{B}} \frac{|\tilde h_0(f)|^2}{S_n(f)}\,(f-\bar{f}_b)\,\Delta f.
\end{align}
Equations~(\ref{eq:Z_banded})--(\ref{eq:hh_banded}), 
together with ~\eqref{eq:likelihood}, provide a fast and accurate approximation to the full log-likelihood using only a small set of band summaries $\{A_0,A_1,B_0,B_1\}$ and the bandwise ratio coefficients $\{\tilde r_{0}(b,\theta),\tilde r_{1}(b,\theta)\}$.
Operationally, $\tilde r_{0}(b,\theta)$ and $\tilde r_{1}(b,\theta)$ are determined at the band edges by dividing the trial waveform by the reference, thereby fixing the intercept and slope per band; 
the same construction yields a corresponding approximation for $\langle h\mid h\rangle$ through ~\eqref{eq:hh_banded}.


The detailed implementation of this approximation is not unique and must be chosen by balancing numerical accuracy against computational cost for the scientific application of interest.
For a signal written in the form of ~\eqref{eq:fd-model}, the likelihood may be constructed either from mode-by-mode template ratios or from channel-by-channel template ratios. One must also specify the number of frequency bins, the expansion order within each bin, and the overall approximation structure. The relevant criterion is whether the resulting likelihood error remains below the statistical uncertainty relevant for parameter inference. In general, finer binning, higher-order expansions, and more structured constructions improve accuracy at the expense of increased computational cost. For more complicated multimode signals, more refined schemes such as mode-by-mode relative binning can offer better numerical error control \cite{Leslie:2021ssu,Narola:2023men}.

In this work, we adopt the computationally cheaper strategy of approximating the total waveform directly in the \ac{TDI} channels and then evaluating the heterodyned likelihood. 
The fidelity of this approximation is mainly determined by the mismatch between the reference template and the target signal, and by the smoothness of the waveform manifold in parameter space near the reference template. We choose the injected waveform $\tilde h_0(f,\theta_0)$ as the reference template, so that no additional parameter offset is introduced between the reference and injected signals. To assess the stability of this local approximation, we draw trial waveforms from the chosen prior distribution and compare the relative error between the heterodyned likelihood and the exact Whittle likelihood~\eqref{eq:likelihood}. For different signals, the required number of frequency bins is ultimately set by the signal properties and by the level of likelihood error tolerable for parameter inference. In the applications considered below, we use a sparse frequency grid consisting of 128 points uniformly spaced in logarithmic frequency, in order to balance rapid sky localization against preservation of the geometry of the parameter space.

\section{Likelihood-free parameter inference}\label{subsec:mbhb-nsf}

\subsection{Basic theory for density estimation}

In many areas of physics, including gravitational-wave astronomy, forward simulators
$x \sim p(x \mid \theta)$ are readily available, whereas an analytic likelihood
$\mathcal{L}(x \mid \theta)$ is intractable or prohibitively expensive to evaluate.
\ac{SBI}
 is designed precisely for this regime. In this work, we use the term \ac{SBI} in the broad sense of inference methods that rely on samples from a forward simulator when the exact likelihood is unavailable or impractical to evaluate directly. In much of the literature, this class of methods is also referred to as likelihood-free inference (LFI). Under this convention, \ac{SBI} includes approaches that learn the posterior directly, as well as methods that learn a surrogate likelihood or likelihood ratio from simulations.
In the present context, the target is the Bayesian posterior distribution defined in ~\eqref{eq:bayes}. By training on simulated signal-plus-noise data $d$ generated from prior samples  $\theta_0 \sim \pi(\theta)$, the model learns an approximate mapping from the joint distribution $p(d,\theta)$ of observations $(d,\theta)$ to the posterior distribution $p(\theta \mid d)$ over parameters. Once trained, it can be applied directly to an observed data stream and return an approximate posterior with low latency.


A variety of \ac{SBI} methods have been developed in the literature, including approximate Bayesian computation, adversarial approaches, score-based methods, and explicit conditional-density estimators~\citep{Marjoram2003,Cranmer_2020PNAS,2021arXiv210104653L,2013arXiv1312.6114K,2015arXiv150505770J,hyvarinen2005estimation}. These classes differ substantially in sample efficiency, training stability, implementation complexity, and posterior expressivity.
For example, distance-based approximate Bayesian computation suffers from rapidly degrading sample
efficiency as dimensionality grows and the acceptance threshold is tightened ~\cite{Marjoram2003,2015arXiv151205633P}; implicit adversarial training can be fragile (e.g., mode collapse)~\citep{2013arXiv1312.6114K}; and score matching methods are highly expressive and multi-modal
friendly but are computationally demanding ~\citep{hyvarinen2005estimation}. 

In this work, we therefore adopt the explicit conditional density route and model the posterior directly as $q(\theta \mid x)$ using the normalizing flow method.
This choice is motivated by the need to represent non-Gaussian and potentially multi-modal posteriors while retaining fast amortized inference after training. Our methodological focus is therefore not only posterior accuracy, but also the construction of a practical low-latency parameter-estimation pipeline.

\subsection{Neural spline flow}





A normalizing flow is a bijective mapping between a target distribution 
$p_{\mathcal{\boldsymbol{X}}}(\boldsymbol{x})$, which we aim to estimate, and a latent distribution 
$p_{{\mathcal{V}}}(\boldsymbol{v})$ that is easy to sample from (e.g., a Gaussian) ~\citep{2019arXiv190809257K}. 
The transformation is realized through a sequence of invertible functions, 
each parameterized by neural networks. Formally, if $\boldsymbol{v} = \mathbf{F}(\boldsymbol{x};\xi)$, 
where $\xi$ denotes the  trainable network parameters, then the probability density of $\boldsymbol{x}$ 
can be expressed as
\begin{equation}
    p_{\mathcal{X}}(\boldsymbol{x}) = p_{\mathcal{V}}(\boldsymbol{v}) \bigg|\frac{\partial \boldsymbol{v}}{\partial \boldsymbol{x}}\bigg|
    = p_{\mathcal{V}}\!\left(\mathbf{F}(\boldsymbol{x};\xi)\right)\,\bigg|\det \frac{\partial (\mathbf{F}(\boldsymbol{x};\xi))}{\partial \boldsymbol{x}}\bigg|. \label{prob_trans}
\end{equation}
The requirement of invertibility guarantees that the transformation $\mathbf{F}(\boldsymbol{x};\xi)$ 
is well-defined in both directions. Efficient computation of the last term in Eq.~(\ref{prob_trans}), the Jacobian 
determinant, ensures that volume changes induced by the transformation are properly accounted for, 
which is crucial for maintaining a valid probability density.

In this work, we employ a coupling \ac{NSF}~\cite{2019arXiv190604032D}, in which each invertible layer is implemented as a coupling transformation~\cite{2016arXiv160508803D} parameterized by  rational quadratic splines (RQ-splines). 
Let a $K$-dimensional input vector be denoted as $x \in \mathbb{R}^K$. 
For the $n$th coupling layer, let the input be split as
$$\mathbf{f}_{n-1}=(\boldsymbol{x}_A,\boldsymbol{x}_B)\in\mathbb{R}^k\times\mathbb{R}^{K-k},$$ 
and let the output be $\mathbf{f}_n=(\boldsymbol{y}_A,\boldsymbol{y}_B)$. 
The coupling map $\mathbf{f}_n$ is defined by
\begin{equation}
\begin{aligned}
    \boldsymbol{y}_A &= \boldsymbol{x}_{A}, \qquad \\
    \boldsymbol{y}_{B,i} &=g_{n,i}\!\left(\boldsymbol{x}_{B,i};\Theta_{n,i}(\boldsymbol{x}_A)\right), \qquad
 i=1,\dots,K-k,
 \end{aligned}
\end{equation}
where the subscript $i$ denotes the $i$-th component of the vectors. Here, $g_{n,i}$ represents a generic one-dimensional RQ-spline transformation applied to $\boldsymbol{x}_{B,i}$. The spline parameters $\boldsymbol{\Theta}_n(\boldsymbol{x}_A) = \{\boldsymbol{\Theta}_{n,i}\}_{i=1}^{K-k}$ are generated by a conditioner network (typically a ResNet) conditioned on the input subspace $\boldsymbol{x}_A$. This architecture ensures that while the functional form of $g$ is shared across all dimensions $i$ and all layers $n$, the transformation parameters are specific to each input. To ensure full interaction between dimensions, the roles of $\boldsymbol{x}_A$ and $\boldsymbol{x}_B$ are swapped in alternating coupling layers.

Under this componentwise construction, the corresponding Jacobian matrix of a single layer takes a lower-triangular form:
\begin{equation}
\mathbf{J}_n
=
\frac{\partial \mathbf{f}_n}{\partial \mathbf{f}_{n-1}^T}
=
\begin{bmatrix}
\mathbb{I}_k & \mathbf{0}_{k\times (K-k)} \\
\dfrac{\partial \boldsymbol{y}_B}{\partial \boldsymbol{x}_A^T} &
\mathrm{diag}\!\left(
\dfrac{\partial g_{n,i}(x_{B,i};\Theta_{n,i}(\boldsymbol{x}_A))}{\partial x_{B,i}}
\right)
\end{bmatrix}.
\end{equation}
whose determinant is the product of its diagonal entries,
\begin{equation}
\det(\boldsymbol{J}_n) = \prod_{i=1}^{K-k} 
\frac{\partial g_{n,i}(\boldsymbol{x}_{B,i}; \Theta_{n,i}(\boldsymbol{x}_A))}{\partial \boldsymbol{x}_{B,i}}.
\end{equation}
and equivalently,
\begin{equation}
\log |\det \mathbf{J}_n|
=
\sum_{i=1}^{K-k}
\log \left|
\frac{\partial g_{n,i}(\boldsymbol{x}_{B,i};\Theta_{n,i}(\boldsymbol{x}_A))}{\partial \boldsymbol{x}_{B,i}}
\right|.
\end{equation}

The full flow is constructed as a composition of $N$ coupling layers,
$
\mathbf{F}=\mathbf{f}_N\circ \mathbf{f}_{N-1}\circ\cdots\circ \mathbf{f}_1.
$
By the chain rule, the Jacobian of the full flow is the product of the layerwise Jacobians,
\begin{equation}
\det\!\left(\frac{\partial \mathbf{F}}{\partial \boldsymbol{x}^T}\right)
=
\prod_{n=1}^{N}\det(\mathbf{J}_n),
\end{equation}
and equivalently,
\begin{equation}
\log \left|
\det\!\left(\frac{\partial \mathbf{F}}{\partial \boldsymbol{x}^T}\right)
\right|
=
\sum_{n=1}^{N}\log |\det(\mathbf{J}_n)|.
\end{equation}

The objective function for training is defined as the Kullback--Leibler divergence 
between the model distribution $q_{\mathcal X}(\boldsymbol{x};\xi)$ and the target distribution $p_{\mathcal X}(\boldsymbol{x})$:
\begin{equation}\label{eq:flow-loss}
\begin{split}
    \mathcal{L}(\xi) =& D_{\textrm{KL}}\!\left[p_{\mathcal X}(\boldsymbol{x})\,\|\,q_{\mathcal X}(\boldsymbol{x};\xi)\right] \\
    =& \int p_{\mathcal X}(\boldsymbol{x})
\ln \frac{p_{\mathcal X}(\boldsymbol{x})}{q_{\mathcal X}(\boldsymbol{x};\xi)}
\,\mathrm{d}\boldsymbol{x} \\
   =& - \int p_{\mathcal X}(\boldsymbol{x})
\ln q_{\mathcal X}(\boldsymbol{x};\xi)
\,\mathrm{d}\boldsymbol{x}
+ \mathrm{constant} \\
=& - \mathbb{E}_{p_{\mathcal X}(\boldsymbol{x})}
\left[\ln q_{\mathcal X}(\boldsymbol{x};\xi)\right]
+ \mathrm{constant}. \\
=&
- \mathbb{E}_{p_{\mathcal X}(\boldsymbol{x})}
\left[
\ln p_{\mathcal{V}}\!\bigl(\mathbf{F}(\boldsymbol{x};\xi)\bigr)
+ 
\ln \left|
\det \frac{\partial \mathbf{F}(\boldsymbol{x};\xi)}{\partial \boldsymbol{x}^T}
\right|
\right]
\\ &
+ \mathrm{constant}.
\end{split}
\end{equation}
Here the constant term depends only on the target distribution and is therefore independent of the model parameters $\xi$. Consequently, minimizing the KL divergence is equivalent to maximizing the log-likelihood of samples drawn from $p_{\mathcal X}$ under the \ac{NSF} model.


In our implementation, the \ac{NSF} model is trained in the normalized space $\boldsymbol{u}=\mathbf{T}(\boldsymbol{x})$ rather than directly in the physical parameter space $\boldsymbol{x}$, as described in the section~\ref{subsubsection:training}.
Because $\mathbf{u}$ is itself obtained from the physical parameters $\boldsymbol{x}$ through the normalization map,
the density in physical space additionally contains the Jacobian of the transformation $\boldsymbol{x}\mapsto\mathbf{u}$.
Accordingly, the empirical loss minimized during training is
\begin{equation}\label{eq:flow-loss}
\begin{split}
 \mathcal{L}(\xi) = & -\frac{1}{M} \sum_{m=1}^{M} 
 \bigg[ \ln p_{\mathcal V}(\mathbf{F}(\mathbf{T}(\boldsymbol{x}_m);\xi)) \\
 &+ 
 \ln \bigg|\det \frac{\partial \mathbf{F}(\mathbf{T}(\boldsymbol{x}_m);\xi)}{\partial \boldsymbol{u}^{T}}\bigg|   
 +   \ln \bigg|\det \frac{\partial \mathbf{T}(\boldsymbol{x}_m)}{\partial \boldsymbol{x}^T } \bigg |  \bigg],
 \end{split}
\end{equation}
where $M$ denotes the number of points in the training space, which are indexed by $m$.


We employ the \texttt{glasflow} package~\cite{michael_j_williams_2024_13914483} to implement the \ac{NSF}. This package uses a ResNet-based architecture to parameterize flexible 
transformations, which are then composed into the overall mapping function $g$. 
In our implementation, we use the \texttt{CouplingNSF} class, defined as a sequence 
of transformation functions acting on the latent variables and conditioned on the 
embedding network output.
In practice, we set \texttt{n\_transforms} $=50$ and choose 
\texttt{n\_conditional\_inputs} to match the dimensionality of the embedding. 
Each transform is built from a ResNet with \texttt{n\_neurons} $=256$ in the hidden 
layers and \texttt{nblocks\_per\_transform} $=32$, with a normalization layer 
applied after each block. The spline discretization uses \texttt{n\_bins} $=10$. 
All remaining hyperparameters controlling the spline shape and coupling structure 
are kept at their default values in \texttt{glasflow}.


\subsection{Embedding }



For posterior estimation, directly feeding the raw strain time series into the \ac{NSF} is inefficient, because the primary role of the flow is to model the target parameter distribution rather than to extract informative features from long sequential data. We therefore introduce an embedding network to compress the detector output into a low-dimensional projection before density estimation. 
In our setup, each sample consists of a  5-day observation segment sampled at 0.01 Hz, corresponding to an input of size $(2,4320)$. This choice is motivated by the low-latency MBHB setting considered here and is broadly consistent with previous space-based deep-learning studies that analyze multi-day pre-merger MBHB data segments on comparable timescales \cite{Ruan:2024qch}. Even at this reduced cadence, the detector output remains a moderately high-dimensional time series whose structure is shaped jointly by the source evolution, the nonlinear TianQin response, and the noise realization~\cite{Zhang:2024fka}.

\begin{table*}[!ht]
\centering
\caption[Embedding Network Structure]{Embedding Network Structure: Concatenating Inception-like module and ResNet. The stride in each layer is 2.}
\label{tab:embeddingnetwork-structure}
\begin{tabular}{|c|c|c|c|}
\hline
\hline
Network Type & Layer Name  & Input Size & Output Size \\
\hline
\multirow{4}{*}{Inception-like module} & 
Branch 1: $8  \times 7$ Convolution, AdaptiveAvgPool(48) &   (2, 4320) & (8, 48) \\
& Branch 2:  $ 8 \times 14$ Convolution, AdaptiveAvgPool(48) &   (2, 4320) & (8, 48) \\
& Branch 3:$8 \times 28$ Convolution, AdaptiveAvgPool(48) &   (2, 4320) & (8, 48) \\
& Branch 4:$8 \times 56$ Convolution, AdaptiveAvgPool(48) &   (2, 4320) & (8, 48) \\
& Branch 5:$8 \times 128$ Convolution, AdaptiveAvgPool(48) &   (2, 4320) & (8, 48) \\
& Branch 6:$1 \times 3$  Maxpool,AdaptiveAvgPool(48)&   (2, 4320) & (2, 48) \\
& Flatten \& Concatenation: Global avgpool & - & (1, 2016) \\
\hline
\multirow{4}{*}{ResNet-50} & 
Cov1:  $96 \times 7$ Convolution, $1 \times 3$ Maxpool &  (1, 2016) &(1, 2304, 504) \\ 
& Multiple Residual Blocks &  (1, 2304, 504) & (1, 2048, 63) \\
& Avgpool & (1, 2048, 63) & (1, 2048) \\
& Fully connected  & (1, 2048) & (1, 128) \\
\hline
\hline
\end{tabular}
\end{table*}

A key requirement for the embedding model is that it must preserve parameter-relevant information distributed across multiple characteristic timescales. 

For MBHB signals in the TianQin band, the waveform evolves over roughly $\mathcal{O}(10^{-4}\text{--}10^{-2})\,\mathrm{Hz}$, while the motion of the TianQin satellite constellation introduces response modulations on longer timescales, with an orbital period of 3.64 days around the Earth ($\sim 3.18 \times 10^{-6}\,\mathrm{Hz}$).  As a result, the input time series contains both relatively long-scale correlated structures and relatively short-scale oscillatory features, spanning roughly two orders of magnitude in characteristic frequency. A single convolution scale is therefore generally insufficient: large receptive fields are needed to capture slowly varying correlations, whereas small receptive fields are needed to resolve local oscillatory structure.

Based on these different time-frequency relation, we adopt an embedding architecture that combines multiple convolution kernels with residual blocks. The final embedded network architecture is summarized in Table \ref{tab:embeddingnetwork-structure}.
In this design, large convolutional kernels, inspired by the Inception framework~\citep{2014arXiv1409.4842S} and the convolutional neural network for detecting extreme mass ratio inspirals signals~\citep{Zhang:2022xuq}, are employed to extract large-scale correlated features corresponding to long-period, low-frequency information. Meanwhile, small convolutional kernels, following the ResNet paradigm~\citep{He:2016cvpr}, are used to refine short-scale feature extraction and improve information propagation through depth. 
The embedding network and the NSF are trained jointly. This design has an important advantage: the compressed representation is learned not as a generic summary of the raw data, but as a task-oriented representation optimized directly for posterior estimation. In other words, the embedding is encouraged to retain precisely those features that are most useful for inferring the source parameters modeled by the \ac{NSF}. This is particularly important here, because the informative content is distributed across multiple temporal scales and a purely linear or separately trained compression stage would be less well matched to the final inference objective. The final embedding network maps the input of size $(2,4320)$ to a 128-dimensional latent representation, which is then passed to the \ac{NSF} for conditional posterior modeling.

\subsection{Training and validation}\label{subsubsection:training}


To achieve good performance with reasonable computational cost, we focus on two key aspects: the selection of training data and the optimisation of the neural-network architecture. 
Compared with data selection, optimizing the network architecture is more urgent. 
Therefore, before settling on a final design, we carried out a series of experiments within a restricted region of parameter space. 
At the end, the prior ranges used in the tests are listed in Table \ref{tab:mbhb_priors}. 
For now, we generate training samples in a nine-dimensional parameter space that does not include spins, i.e.\ the component spins are fixed to zero.
The \ac{NSF} model is then trained to estimate the posterior distribution for six parameters: the chirp mass, ecliptic longitude, ecliptic latitude, symmetric mass ratio, coalescence time, and luminosity distance. 
The remaining three parameters are randomized over their prior distributions during training and are therefore treated in a marginalized sense.

Before training, we map the physical parameters $\boldsymbol{\theta}$ to normalized coordinates $\mathbf{u}\in(0,1)^K$ using parameter-dependent monotonic transformations. This normalization places parameters on comparable numerical scales, preserves physical bounds and angular periodicity, and improves training stability.
For each $\theta_i\in[\theta_{i,\min},\theta_{i,\max}]$, we define $u_i=T_i(\theta_i)$. Most parameters are mapped linearly,
\begin{equation}
u_i=\frac{\theta_i-\theta_{i,\min}}{\theta_{i,\max}-\theta_{i,\min}}.
\end{equation}
For the chirp mass $\mathcal{M}_c$, we use
\begin{equation}
u_{\mathcal{M}_c}
=
\frac{\log_{10}\mathcal{M}_c-\log_{10}\mathcal{M}_{c,\min}}
{\log_{10}\mathcal{M}_{c,\max}-\log_{10}\mathcal{M}_{c,\min}},
\end{equation}
for the latitude $\lambda$,
\begin{equation}
u_{\lambda}=\frac{\lambda \bmod 2\pi}{2\pi},
\end{equation}
and for the luminosity distance $D_L$,
\begin{equation}
u_{D_L}
=
\frac{D_L^{\,3}-D_{L,\min}^{\,3}}
{D_{L,\max}^{\,3}-D_{L,\min}^{\,3}}.
\end{equation}
Here the logarithmic, wrapped-angular, and power-law maps are chosen to accommodate the broad dynamic range of $\mathcal{M}_c$, the periodicity of $\lambda$, and the nonlinear scaling of $D_L$.

For the network design, we initially explored a ResNet-based embedding network that was first pretrained on a parameter point-estimation task, after which the corresponding compressed representation was passed to the \ac{NSF}. The motivation for this strategy was to preserve as much source-parameter information as possible during dimensionality reduction while avoiding the introduction of an additional decoder, thereby reducing both training cost and memory usage. 
However, this approach proved inefficient, requiring more than 10 days of training, and was practically incapable of achieving lossless projection across all source signals.  

Upon closer analysis, we observed that some source parameters exhibit strong degeneracies. For example, sky localization is degenerate due to the intrinsic 
characteristics of the detector (e.g., the fixed-plane orientation of the TianQin mission), where different sky locations can produce nearly identical responses in 
the detector frame. Consequently, when trained solely on the point estimation task, the embedding network struggles to distinguish signals with similar inputs but different parameter values, leading to nearly identical compressed representations for distinct parameter sets. 
This effect has been demonstrated in the likelihood analyses of \ac{MBHB} inspirals with different signal durations by Marsat 
et al.~\cite{Marsat:2020rtl}.


%
Given this, and the desire to provide early warning for EM follow-up, we place particular emphasis on minimizing errors in sky localization. 
we ultimately adopted a stage-wise hyperparameter-tuning and training strategy in which the embedding network and the \ac{NSF} were trained jointly. 
We first optimize the flow architecture, then refine the embedding network, and finally reassess the stability of the flow under the full inference setting. 
In Stage~I, we restrict the target to the sky-localization parameters and identify the best-performing NSF configuration. In Stage~II, we gradually expand the inferred parameter set to the six key source parameters, while progressively optimizing the embedding architecture. In Stage~III, we jointly fine-tune all model components using $10^{7}$ training samples and $5\times10^{5}$ validation samples, yielding an intermediate final model.. A full training run at this stage required 223 hours on an NVIDIA A100 GPU with 80~GB of memory, using a batch size of 2048. Under this setup, the dominant computational bottleneck arises from the forward or backward pass of the embedding model. 
Starting from this model, we then performed an additional retraining stage (Stage~IV), in which the initial learning rate was reduced from $10^{-4}$ to $10^{-5}$ and the model was further optimized using $10^{7}$ newly generated training samples drawn from the same prior distribution, together with the same $5\times10^{5}$ validation samples as in Stage~III. This final retraining stage required 217 hours.

\begin{figure}[htbp]
\centering
\includegraphics[width=0.48\textwidth]{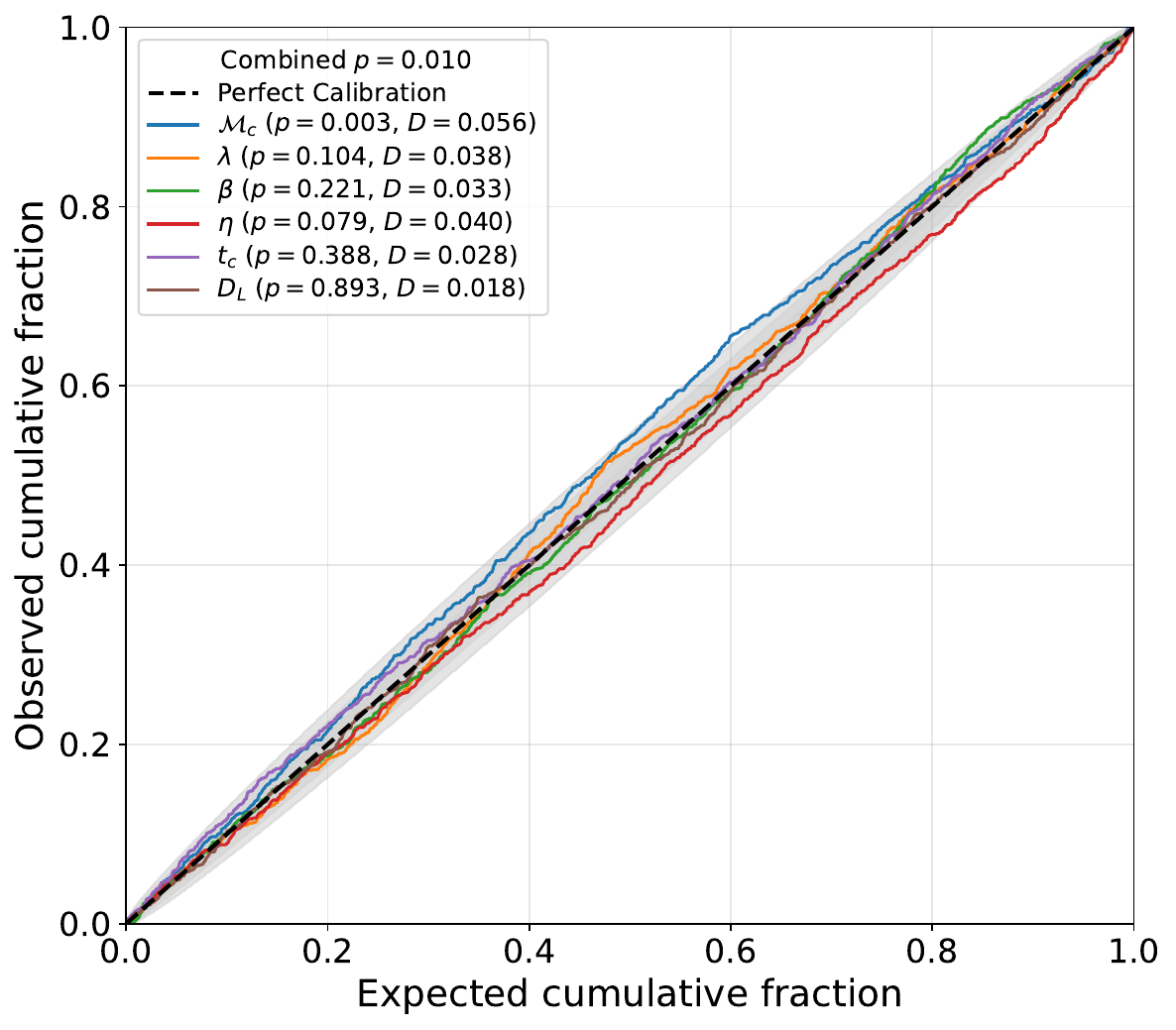}
\caption{
P--P plot assessing the calibration of the NSF posteriors over 1000 events. 
The dashed black line indicates perfect calibration, and the shaded bands denote 
the $1\sigma$, $2\sigma$, and $3\sigma$ binomial credible regions. The legend displays the Kolmogorov-Smirnov (K-S) test results for each parameter, 
listing the \textit{p}-value first to indicate statistical consistency with a uniform distribution, 
followed by the K-S statistic ($D$) which quantifies the maximum absolute deviation from the diagonal.}
\label{fig:pp_plot}
\end{figure}

Unless otherwise specified, all models are trained for up to 200 epochs using the Adam optimizer with an initial learning rate of $1 \times 10^{-4}$. Cosine learning rate annealing is applied to schedule the learning rate, and early stopping is implemented with a patience of 30 epochs (i.e., training terminates if the validation loss fails to improve for 30 consecutive epochs) to prevent overfitting.
As shown in Fig.~\ref{fig:pp_plot}, 
the best-trained model yields Probability-Probability (P--P) plots 
 where the empirical cumulative distributions closely follow the diagonal line of perfect calibration. 
Most physical parameters (e.g., $\lambda$, and $\beta$) exhibit high consistency with the uniform distribution, characterized by \textit{p}-values greater than 0.05 and small K-S statistics ($D < 0.04$). This indicates that the posterior estimates for these parameters are self-consistent and statistically unbiased.
Although the chirp mass ($\mathcal{M}_c$) yields a lower \textit{p}-value of 0.003, the corresponding K-S statistic is $D=0.056$. This implies that the maximum absolute deviation from the ideal calibration is only $5.6\%$. 
Finally, the combined \textit{p}-value for the global test is 0.010, further supporting the overall reliability of the model calibration across all parameters.

\section{Results}\label{subsec:mbhb-results}
We will now describe our results, demonstrating the performance of the NSF network and benchmarking it against standard sampling techniques.



\subsection{Sampler accuracy and latency }\label{sec:acc_latency}

To assess the practical performance of our inference pipeline, we first compare 
the trained NSF-based sampler against a reference \ac{PTMCMC} analysis for a single, representative 
MBHB injection.
The injected source parameters are listed in Table~\ref{tab:injection-params}. 
We consider a non-spinning binary with redshifted chirp mass 
$\mathcal{M}_c = 9\times10^{5}\,M_{\odot}$ at redshift $z=2$ 
(corresponding to $D_L \approx 16\,\mathrm{Gpc}$), inclined at 
$\iota = \pi/3$ and merging  $\Delta t_c = 15\,\mathrm{min}$ after the end of the observation. 
Both samplers are run on the same injection and prior. \ac{PTMCMC} serves as a high-accuracy baseline, allowing the NSF to be assessed in terms of both accuracy and 
latency.

\begin{table}[htbp]
\centering
\caption{Injection parameters for the test signal. The black holes are assumed to be non-spinning ($\chi_{1z}=0, \chi_{2z}=0$).
The observation window spans 5 days with $t_0=0~ $s defined as the start time. The coalescence time is set to  $t_c=432900 $s relative to $t_0$ , such that the merger occurs 15 minutes after the end of the observation.
}
\label{tab:injection-params}
\begin{tabular}{ll}
\toprule
\textbf{Parameter} & \textbf{Value} \\
\midrule
 Chirp mass $\mathcal{M}_c$ & $9 \times 10^{5}\ M_{\odot}$ \\
Symmetric mass ratio $\eta$ & $0.2$ \\
Redshift $z$ & $2$ \ (corresponding to $D_L \approx 16\,\mathrm{Gpc}$) \\
Orbital inclination $\iota$ & $\pi/6\ \mathrm{rad}$ \\
Coalescence phase $\phi_c$ & $\pi/2\ \mathrm{rad}$ \\
Ecliptic longitude $\lambda$ & $\pi/3 \ \mathrm{rad}$ \\
Ecliptic latitude $\beta$ & $\pi/3\ \mathrm{rad}$ \\
Coalescence time $ t_c$ &  $432900 $ seconds  \\
Polarization angle $\psi$ & $\pi/3\ \mathrm{rad}$ 
\\ 
\ac{SNR} & 162.9  \\
\bottomrule
\end{tabular}
\end{table}

\begin{figure*}[htbp]
\centering
\includegraphics[width=0.9\textwidth]
{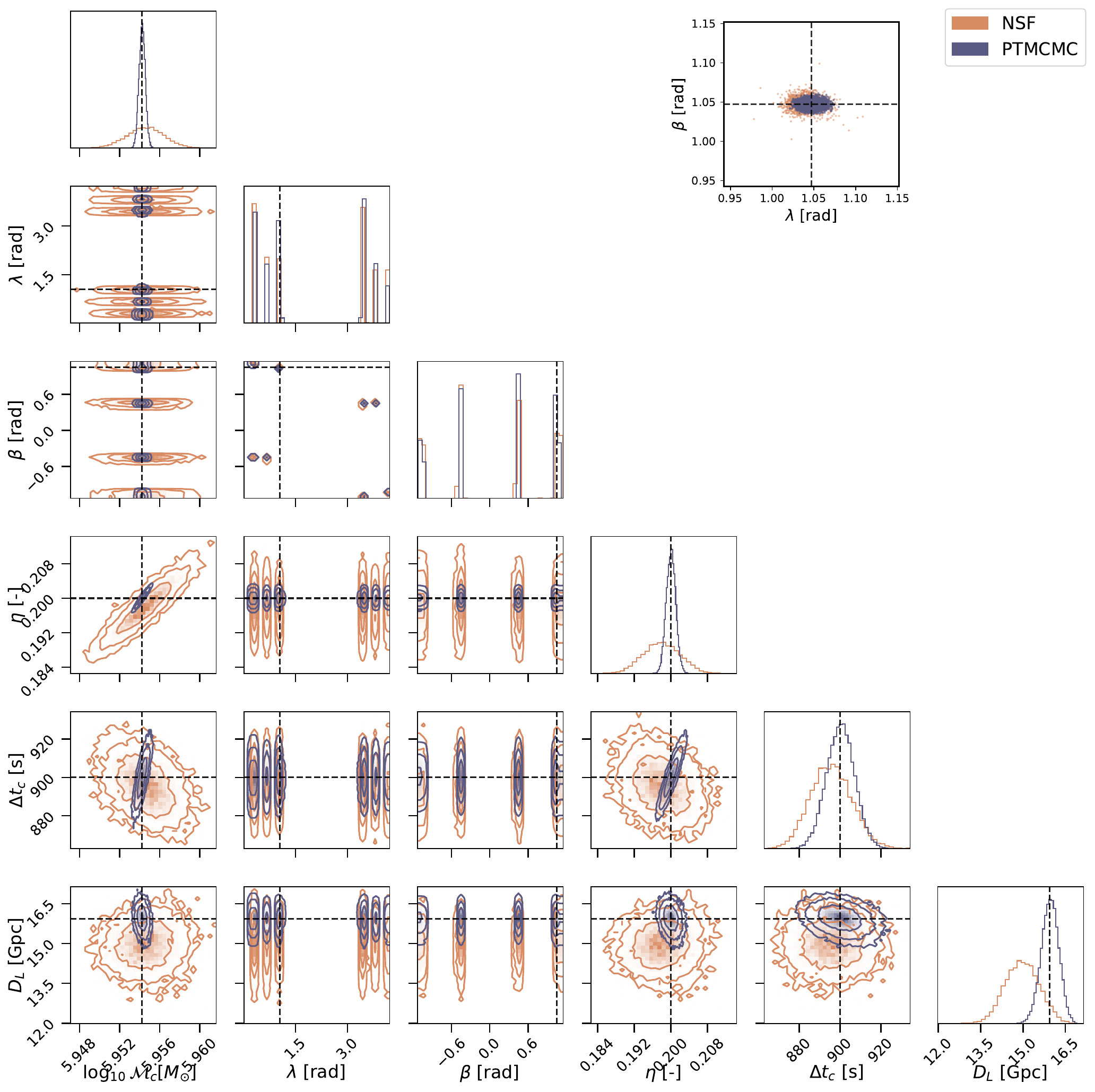}
\caption{
Comparison of posterior distributions for a representative MBHB injection: \ac{PTMCMC} (slate blue) computed on noise-free data versus trained NSF (terracotta) evaluated on noisy data. 
The NSF samples represent unfiltered raw outputs without rejection of discrete outliers. 
Contours indicate joint $3\sigma$ credible regions, with dark cross-hairs marking the injected parameter values.
}
\label{fig:all-params-prediction}
\end{figure*}

Figure~\ref{fig:all-params-prediction} compares posterior samples obtained with our NSF inference pipeline and a reference \ac{PTMCMC} analysis for the same event. The plot overlays the one-dimensional marginal distributions and the two-dimensional joint credible contours for the six inferred parameters
$\{\log \mathcal{M}_c, \lambda, \beta, \eta, \Delta t_c, D_L\}$. The remaining three parameters, $(\iota,\phi_c,\psi)$, are not sampled explicitly by the NSF; instead, they are randomized over their priors during training so that their effects are learned in a marginalised sense. 
The injected (true) parameter values are marked by the dark lines. The two samplers recover consistent high-posterior regions in the well-constrained dimensions, and the injected truth lies within the dominant credible regions. 

To characterize the statistical properties of the posterior samples more completely, in addition to reporting the $1\sigma$ credible widths of the one-dimensional parameters, we also provide the sky-localization widths and sky area at the $90\%$ credibility level.
For the sky-position parameters, the marginal widths $\Delta\lambda$ and $\Delta\beta$ alone are not sufficient to fully capture the two-dimensional posterior structure, especially when the sky posterior exhibits clear multimodality. In such cases, the sky area provides a more direct measure of the overall geometric extent of the localization region. In this work, the sky area is computed using a HEALPix-based pixelization scheme~\cite{Gorski:2004by}, with implementation details given in Appendix~\ref{appendix:c}. The corresponding parameter-precision, divergence, and sky-localization results are summarized in Tables~\ref{tab:mbhb_pe_precision},  \ref{tab:mbhb_sky_stats} and \ref{tab:kl_js_1d_main_params}.

For a general multimodal sky posterior, we use a clustered area estimate: distinct sky modes are first identified, and the credible sky area of each mode is defined as the area of its \ac{HPD} region at the chosen credibility level; the total sky area is then obtained by summing over all modes. 
In what follows, such quantities are denoted by the subscript ``clustered''. For the TianQin detector, whose sky response exhibits discrete symmetries, multiple sky modes in the posterior can under suitable conditions be interpreted as symmetry-equivalent images of the same physical solution. Motivated by this property, we further introduce mode folding as an alternative and independent characterization of sky localization, in which symmetry-related samples are folded into a single representative mode, from which the folded longitude width, latitude width, and sky area are computed.
The former describes the full multimodal sky-localization region, whereas the latter characterizes the effective single-mode localization scale after removing the symmetry-related degeneracy. The two should therefore be regarded as complementary rather than interchangeable localization metrics.

Overall, both methods yield informative and astrophysically useful posterior constraints, but with a clear division between the parameters most relevant for early warning and those that probe finer intrinsic structure.
In particular, the reported fractional uncertainties remain below $10\%$ for both samplers. 
For the quantities that matter most for rapid alert generation, namely sky localization and arrival time, NSF remains comparable in scale to the \ac{PTMCMC} reference. At the $90\%$ credibility level, the single representative sky-mode widths are $\Delta\lambda_{\rm m}=0.0287$ and $\Delta\beta_{\rm m}=0.0292$ for NSF, compared with $0.0265$ and $0.0131$ for \ac{PTMCMC}, while the corresponding representative sky areas are $3.34~\mathrm{deg}^2$ and $0.79~\mathrm{deg}^2$, respectively. Likewise, the timing uncertainty is $20.2\,\mathrm{s}$ for NSF and $13.6\,\mathrm{s}$ for \ac{PTMCMC}, both of which are small compared to the $\sim 15\,\mathrm{min}$ interval between the end of the analyzed data segment and merger. This shows that, although NSF is somewhat less precise, it still retains the level of timing and directional accuracy needed for low-latency early-warning applications.

By contrast, the larger differences appear in the best-constrained intrinsic parameters, especially the chirp mass and symmetric mass ratio. The $1\sigma$ widths in Table~\ref{tab:mbhb_pe_precision} show that NSF is broader by factors of about $6$ in $\log \mathcal{M}_c$ ($0.0037$ versus $0.0006$) and $4$ in $\eta$ ($0.0082$ versus $0.0020$). This trend is consistent with the divergence measures in Table~\ref{tab:kl_js_1d_main_params}: both samplers extract substantial information relative to the prior, but the strongest information gain is obtained for $\mathcal{M}_c$, with $\mathrm{KL}=4.57$ for NSF and $6.37$ for \ac{PTMCMC}, whereas the weakest is found for $D_L$, with $\mathrm{KL}=1.01$ and $1.35$, respectively. The same table also shows that the two samplers agree most closely for $D_L$ and $t_c$, where the inter-sampler divergences are very small ($\mathrm{JSD}=0.043$ and $0.027$), but differ more noticeably for the mass-related parameters, especially $\mathcal{M}_c$ ($\mathrm{JSD}=0.362$) and $\eta$ ($\mathrm{JSD}=0.274$). Physically, this pattern suggests that NSF captures the broad posterior structure and the parameters governing early-warning localization reliably, while tending to smooth over part of the finer information content that most strongly constrains the intrinsic binary parameters.
This behavior is also consistent with the expectation that the embedding network, while highly effective for rapid inference, may discard part of the fine-scale information carried by the signal, particularly information associated with higher-frequency waveform structure.


\begin{table*}[!ht]
\caption{Comparison of parameter-estimation precision between \ac{NSF} and \ac{PTMCMC}, computed from the posteriors shown in Fig.~\ref{fig:all-params-prediction}. We report both the absolute $1\sigma$ credible interval widths and the corresponding relative errors.}
\centering
\begin{tabular}{lcccccccc}
\toprule
\multirow{2}{*}{Sampler}
 & $\Delta \log M_c$   
 & $\frac{\Delta (\log M_c)}{ \log M_c}$ 
 & $\Delta \eta$  
& $\frac{\Delta \eta}{ \eta}$ 
 & $\Delta t$ 
 & $\frac{\Delta t}{\Delta t_c}$
 & $\Delta D_L$
 & $\frac{\Delta D_L}{D_L}$ 
\\
& [$M_{\odot}$]
& [\%]
& [$-$]
& [\%]
& [s]
& [\%]
& [Gpc]
& [\%] 
\\
\midrule
NSF     
& 0.0037 & 0.0616
& 0.0082 & 4.0881 
& 20.1875  &2.2431
& 1.2281 & 7.7070 
\\
\ac{PTMCMC}  
& 0.0006 & 0.0102 
& 0.0020 & 1.0183
& 13.6224 &1.5136
& 0.6167  &3.8704 
\\
\bottomrule
\end{tabular}
\label{tab:mbhb_pe_precision}
\end{table*}


\begin{table}[!ht]
\caption{Comparison of sky-localization statistics between the NSF and \ac{PTMCMC} samplers for the same event. Here $N_{\rm mode}$ is the number of sky modes identified by clustering, $\hat A_{\rm tot,clustered}$ is the total sky area obtained by summing over all clustered modes, and $(\Delta\lambda_{\rm m}, \Delta\beta_{\rm m}, \hat A_{\rm m})$ characterize the folded representative mode after removing the discrete symmetry-related degeneracy of the TianQin response. All sky-localization quantities in this table are quoted at the $90\%$ credibility level.}
\centering
\begin{tabular}{lccccc}
\toprule
Sampler
& $N_{\rm mode}$
& $\hat A_{\rm tot,clustered}$
& $\Delta\lambda_{\rm m}$
& $\Delta\beta_{\rm m}$
& $\hat A_{\rm m}$ \\
& [$-$]
& [deg$^2$]
& [rad]
& [rad]
& [deg$^2$] \\
\midrule
NSF
&8
& 20.3790
& 0.0287 & 0.0292
& 3.3440\\
\ac{PTMCMC}
& 8
& 5.2849
& 0.0265 & 0.0131
& 0.7868 \\
\bottomrule
\end{tabular}
\label{tab:mbhb_sky_stats}
\end{table}

\begin{table}[!ht]
\caption{One-dimensional divergence metrics for key parameters.  
KL divergence and JS divergence quantify posterior-prior information gain (columns 2--5) and inter-sampler consistency (columns 6--7).}
\centering
\begin{tabular}{lcccccc}
\toprule
Parameter & \multicolumn{2}{c}{NSF} & \multicolumn{2}{c}{PTMCMC} & \multicolumn{2}{c}{PTMCMC vs NSF} \\
\cmidrule(lr){2-3} \cmidrule(lr){4-5} \cmidrule(lr){6-7}
& KL & JS & KL & JS & KL & JS \\
\midrule
$\mathcal{M}_c$ 
& 4.567 & 0.656 
& 6.367 & 0.684 
& 1.416 & 0.362 \\
$\lambda$  
& 1.587 & 0.389
& 1.669 & 0.402
& 0.044 & 0.011 \\
$\beta$
&1.588 & 0.396
&1.639 & 0.403
&0.053 & 0.013 \\
$\eta$ 
& 2.667 & 0.546 
& 3.874 & 0.631 
& 1.009 & 0.274 \\
$D_L$ 
& 1.014 & 0.283 
& 1.347 & 0.350 
& 0.152 & 0.043 \\
$t_c$ 
& 2.763 & 0.556 
& 3.041 & 0.581 
& 0.093 & 0.027 \\
\bottomrule
\end{tabular}
\label{tab:kl_js_1d_main_params}
\end{table}


Beyond accuracy, latency is the key requirement for early-warning applications. 
In our current implementation, the NSF generates $10^4$ posterior samples for a single event in less than $10\,\mathrm{s}$ on a GPU and in about $1\,\mathrm{min}$ on a CPU, enabling near real-time parameter inference. By contrast, the reference \ac{PTMCMC} analysis is substantially more time-consuming. To achieve convergence, \ac{PTMCMC} generally requires many more sampling steps, and its total runtime includes not only the subsequent heterodyned-likelihood sampling but also the search for a suitable reference template. 
For the representative configuration here with 10 temperature levels, 64 parallel chains, and 500000 sampling iterations, the heterodyned-likelihood sampling converges in approximately 4 hours on a computing node with 56 CPU cores (Intel Xeon 6330 processor).
Moreover, this estimate is optimistic, because in this example we directly set the reference template to the injected signal and therefore did not include either the $\sim 30\,\mathrm{minutes}$ typically required to identify a suitable reference template or any additional parameter-estimation error introduced by template mismatch. Although the exact runtime depends on the implementation and the event properties, this comparison clearly demonstrates the substantial practical latency advantage of the NSF for rapid early-warning inference. Additional representative examples are provided in Appendix~\ref{appendix:a}, and they support the same overall conclusion regarding both computational cost and relative estimation precision.

This latency advantage has direct scientific value because it makes it possible to analyze data segments much closer to merger. For the representative event considered here, the merger takes place about $15\,\mathrm{min}$ after the end of the analyzed observation segment. An NSF latency of $\mathcal{O}(1\,\mathrm{min})$ on a CPU therefore still leaves sufficient time to distribute an alert and initiate \ac{EM} follow-up observations while the binary remains in the inspiral phase. By contrast, \ac{PTMCMC}-type samplers are generally too slow for such near-merger low-latency analyses, even though these late-inspiral data segments are precisely the ones that provide the strongest sky-localization constraints.


\subsection{\ac{EM} follow-up readiness and robustness}\label{sec:em_followup}
\begin{figure*}[htbp]
\centering
\includegraphics[width=1\textwidth]
{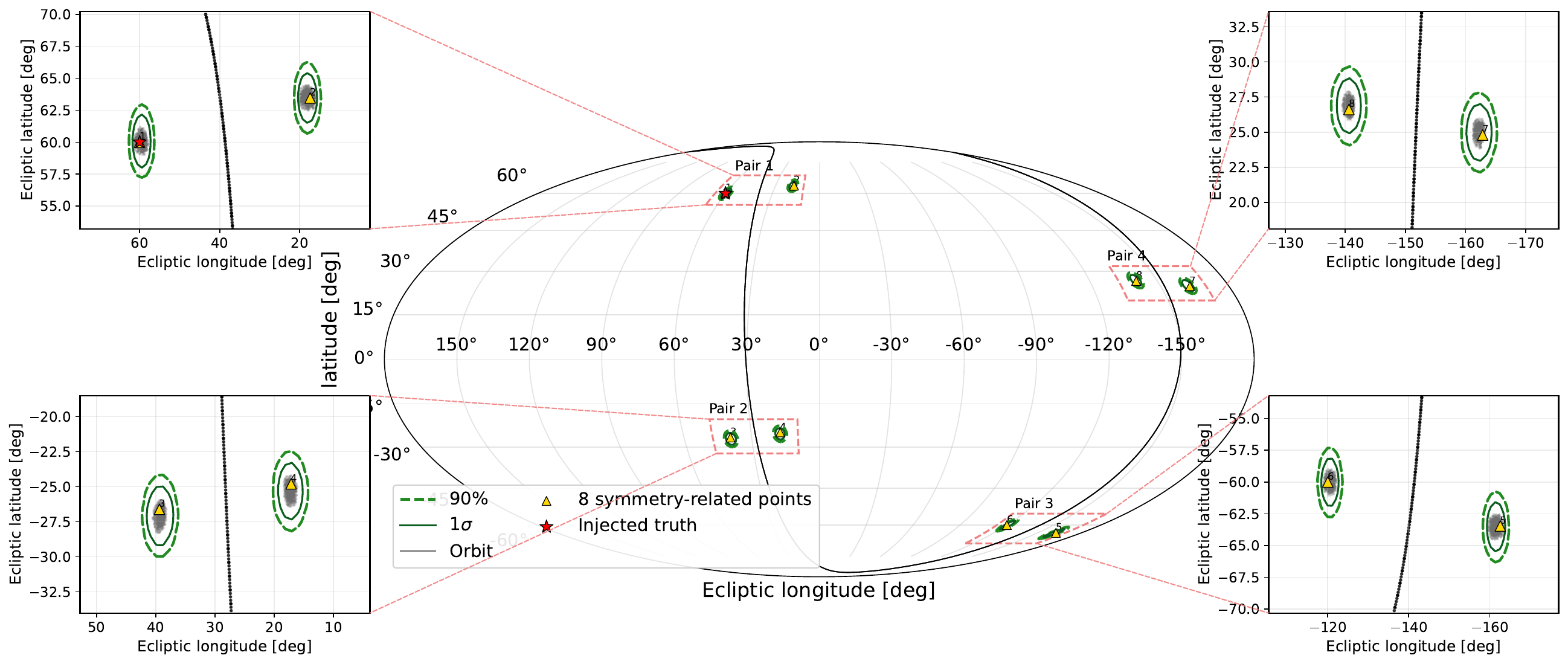}
\caption{
NSF sky map for the representative MBHB event in ecliptic coordinates. 
The yellow triangles mark theoretical symmetry-related sky locations induced by the TianQin detector configuration, and the black points show the TianQin's orbit. }
\label{fig:skymap}
\end{figure*}

We quantify sky-localization performance using the 90\% credible sky area, $A^{90}$. 
For the same representative event considered in the previous subsection, the NSF  achieves an early-warning localization of $A^{90} \approx 20~\mathrm{deg}^2$ using 10000 posterior samples. 
The sky area is evaluated from the HEALPix~\cite{Gorski:2004by} representation of the posterior samples and the corresponding \ac{HPD} region. The resulting sky map is shown in Figure ~\ref{fig:skymap}.


To assess follow-up feasibility, we decompose the end-to-end response time as
\begin{equation}\label{eq:latency_budget}
\begin{split}
t_{\rm total} \approx & t_{\rm inf}+t_{\rm alert}+t_{\rm follow}, \\
=&  t_{\rm inf}+t_{\rm alert}+N_{\rm tile}\,(t_{\rm exp}+t_{\rm read}) +t_{\rm oh},
\end{split}
\end{equation}
where $t_{\rm inf}$ is the inference time, $t_{\rm alert}$ is the alert-distribution latency (including general coordinates network dissemination and basic facility-level triggering), and $t_{\rm exp}$ and $t_{\rm read}$ denote the exposure time and readout time per pointing, respectively.  
Here $t_{\rm oh}$ represents the effective overhead, such as repointing time between adjacent tiles and filter changes. 
The number of pointings required to tile the high-probability localization region is
\begin{equation}
N_{\rm tile}= \max\left(\left\lceil \frac{A^{90}}{\Omega_{\rm FoV}} \right\rceil, N_{\text{mode} }\right),
\label{eq:ntile}
\end{equation}
with $\Omega_{\rm FoV}$ the instrument field-of-view (FoV).

Table~\ref{tab:followup} lists illustrative FoVs, readout times (when available), and simple tiling-time estimates for representative facilities.
These estimates should be interpreted as lower limits, since they neglect additional operational overheads, represented here by $t_{\rm oh}$, and assume idealized tiling under favorable observing conditions. In particular, they do not account for weather losses, visibility constraints, scheduling interruptions, or the possibility that some disconnected localization modes may be simultaneously unobservable from a given site. The estimates therefore apply only insofar as the relevant tiles are observable. With this caveat, 
 the table still shows that facilities such as Rubin/LSST and ZTF can plausibly cover $A^{90}\approx 20 ~\mathrm{deg}^2$ within minutes, consistent with rapid-response follow-up workflows.

For the alert-distribution latency $t_{\rm alert}$, we follow the performance of the LIGO--Virgo--KAGRA public-alert infrastructure. 
\ac{GW} alerts are distributed via the general coordinates network as machine-readable notices and human-readable circulars. 
In particular, the \texttt{LVC\_PRELIMINARY} notice, which includes an automatically generated sky map, is typically issued 
within $1$--$10\,\mathrm{min}$ of the trigger time, while pre-merger \texttt{LVC\_EARLY\_WARNING} notices can be sent with latencies of order tens of seconds or even slightly before merger \citep{GCN_LVK_Mission,LVK_Alerts_UserGuide,Chaudhary:2023vec}. 
We therefore adopt a fiducial $t_{\rm alert}\sim 1\,\mathrm{min}$ in our latency budget, with a conservative upper range of $t_{\rm alert}\lesssim 10\,\mathrm{min}$ to account for variations in event properties and network conditions.

With $t_{\rm inf} \sim 1\,\mathrm{min}$ for the NSF analysis and 
$t_{\rm alert} \sim 1\,\mathrm{min}$ for rapid alert dissemination, 
\ac{EM} facilities can begin responding within a few minutes of the GW trigger. 
For $A^{90} \approx 20~\mathrm{deg}^2$, modern wide-field optical survey instruments can cover the $90\%$ credible region with a small number of pointings, as illustrated in Table~\ref{tab:followup}. 
In the $\sim 10\,\mathrm{min}$ early-warning scenario considered here, this leaves a non-negligible margin to obtain prompt optical observations well before merger.
These simple estimates show that, for the representative event considered here, the NSF-based early warning is fast enough that optical facilities can realistically target even short-lived, high-energy flares occurring in the final minutes before coalescence. 

\begin{table*}[t]
\centering
\caption{Illustrative tiling times for a $90\%$ credible sky area $A^{90}=20~\mathrm{deg}^2$ with 8 modes}
\label{tab:followup}
\begin{tabular}{lcccccl}
\hline
Facility & FoV [deg$^2$] & $t_{\rm read}$ [s] & $t_{\rm exp}$ [s] &
$N_{\rm tile}$ & Imaging time [min] & Reference \\
\hline
Rubin/LSST & 9.6 & 2  & 30 & 8 &
$\approx 8 \times(30+2)/60 \approx 4.3 $ & \cite{2009arXiv0912.0201L} \\
ZTF 
& 47 & 10 & 30 & 8 &
$\approx  8 \times (30+10)/60 \approx 5.3 $ & \cite{2019PASP..131a8002B} \\
Pan-STARRS1 & 7.068 & --- & 45 - 60 & 8 &
$\approx 8 \times 60 / 60 = 8 $ & \cite{2010SPIE.7733E..0EK, 2016arXiv161205560C} \\
 {Athena} & 0.44 & --- & $10^4-10^5$  & 46 &
 $\approx 46 \times(10^5)/60=76.7$ & \cite{Nandra:2013jka,LISA:2022yao} \\
\hline
\end{tabular}
\end{table*}

To test the robustness of this conclusion against noise fluctuations, we further examined ten independent noise realizations for the same injected signal. As summarized in Table~\ref{tab:followup_noise}, the recovered sky area remains broadly stable across these realizations, with a mean value of order $\sim 20~\mathrm{deg}^2$, as the inferred number of sky modes is likewise unchanged. As a result, the corresponding end-to-end follow-up time for most of the \ac{EM} telescopes remains the same across different noise realizations. This indicates that, for a fixed signal, the NSF sampling and sky-area estimation are robust to noise realizations.


\begin{table}[t]
\centering
\caption{Sky localization for 10 independent noise realizations of the same injected signal. For each realization, we report the recovered $90\%$ credible sky area $A^{90}$. Since the number of predicted sky modes is consistently 8 across all noise realizations, the number of disconnected sky modes $N_{\rm tile}$ and the corresponding estimated follow-up time budget $t_{\rm follow}$ remain identical to those reported in Table~\ref{tab:followup}.}
\label{tab:followup_noise}
\begin{tabular}{cc}
\hline
Noise realization & $A^{90}$ [deg$^2$] \\
\hline
1 & \texttt{19.75} \\
2 & \texttt{20.64} \\
3 & \texttt{20.33} \\
4 & \texttt{19.67} \\
5 & \texttt{21.95} \\
6 & \texttt{19.26} \\
7 & \texttt{20.23} \\
8 & \texttt{20.38} \\
9 & \texttt{20.56} \\
10 & \texttt{19.64} \\
\hline
Mean $\pm$ std. & \texttt{20.243 $\pm$ 0.714} \\
\hline
\end{tabular}
\end{table}

\begin{figure}[htbp]
\centering
\includegraphics[width=0.45\textwidth]{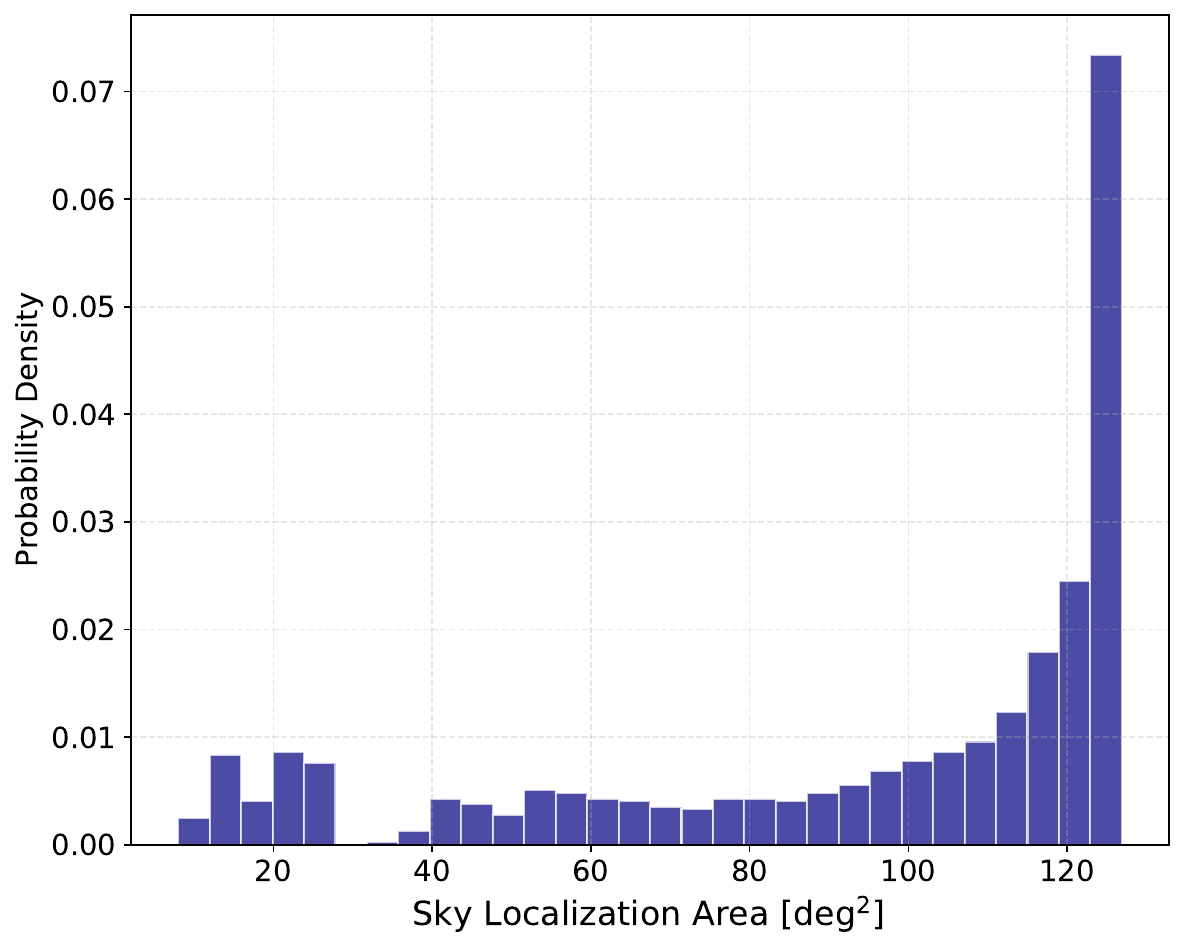}
\caption{Histogram of the recovered 90\% sky-localization areas for 1000 simulated signals. The source parameters are sampled from the prior distribution, and each signal is injected into an independent random noise realization.}
\label{fig:sky_hist}
\end{figure}

More generally, the feasibility of rapid EM follow-up is not uniform across the parameter space. To characterize this variation, we examined the sky-localization performance over 1000 simulated signals. The resulting histogram~(see Fig.~\ref{fig:sky_hist}) indicates that, in addition to the three representative events highlighted in this work, a broader subset of sources achieves recovered localization areas below $100~\mathrm{deg}^2$. This suggests that the scientific utility of a fast NSF-based pipeline is not confined to the specific case studies shown here, but extends to a non-negligible class of comparatively well-localized events. Because the NSF inference latency is expected to remain relatively stable for single-event analyses, these sources are among the most promising candidates for benefiting from rapid inference in support of pre-merger EM follow-up. 
By contrast, if the clustered sky localization is broad or fragmented into many disconnected modes, the event is localized less precisely, and the sky area requiring coverage by \ac{EM} facilities before merger becomes significantly larger. Targeted pre-merger follow-up is therefore less practical in such cases. Such events are instead better suited to archival searches or offline analyses.



\section{Conclusion and outlook}\label{sec:conclusions}


In this work, we have developed a normalizing flow-based \ac{GW} inference pipeline for MBHBs in a TianQin-like configuration, with the explicit goal of enabling early-warning support for short-lived, high-energy \ac{EM} flares. 
The pipeline combines a learned embedding network with a \ac{NSF} posterior estimation model, and is trained to map detector data directly to posterior samples for the main \ac{MBHB} parameters and marginalize the rest of the parameters.

Our results show that the \ac{NSF} reproduces the posterior structure and parameter uncertainties obtained with a reference \ac{PTMCMC} analysis. 
For a representative \ac{MBHB} at cosmological distance, the \ac{NSF} achieves sky localization, distance, and coalescence-time constraints that are consistent with \ac{PTMCMC} to within factors of a few, while reducing the time-to-posterior from tens of minutes to order one minute. 
In particular, incorporating additional signal information from the last hour before merger can significantly improve sky localization, since the localization precision scales approximately with the inverse square of the \ac{SNR}~\cite{Wang:2019ryf}. Previous studies~\cite{Feng:2019wgq} found that, for the target MBHB sources considered here, the signal accumulated during the final hour before coalescence contributes about 90\% of the full-signal \ac{SNR}. This indicates that most of the localization information is concentrated near merger. Extending the analysis further into this final hour can therefore substantially reduce the sky area, but at the cost of a shorter warning window.
This trade-off is challenging for \ac{PTMCMC}-based methods. Because \ac{PTMCMC} inference typically requires minutes to hours, it is difficult to incorporate additional late-time high-\ac{SNR} information while still issuing timely alerts. As a result, conventional approaches often rely on earlier, relatively lower-\ac{SNR} data for preliminary localization, leading to substantially larger sky areas. The low latency of the NSF alleviates this limitation.
In this regime, pre-merger sky areas of order $ \mathcal{O}(10)   \ \mathrm{deg}^2$ can be reached several minutes before coalescence, which is sufficient for wide-field optical facilities to tile the localization region and begin follow-up before merger. 
Diagnostics with multiple noise realizations and different signal scenarios
indicate that the \ac{NSF} posteriors are stable under noise realizations and closely track the \ac{PTMCMC} baseline, with the main discrepancies confined to very narrow, mass-related directions where the true posterior occupies an extremely thin manifold.

The broader multi-messenger problem, however, is more demanding than a single benchmark event. 
The relevant MBHB population spans a wide range in mass, redshift, inclination, and spin, and the associated \ac{EM} counterparts cover multiple timescales and emission channels. 
It is therefore unrealistic to expect a single, static pipeline to solve the entire end-to-end problem in one step. 
Instead, the present work should be viewed as a building block: it demonstrates that fast, calibrated flow-based inference can meet the latency and precision requirements for triggering follow-up of near-merge \ac{EM} counterparts, and can provide a compact probabilistic summary that is suitable for downstream use.

Several extensions follow naturally from this point. 
On the inference side, there is clear room to improve both accuracy and generalizability. 
Firstly, applying the method to longer observation windows and to larger, astrophysically motivated MBHB populations will allow us to assess performance across the relevant parameter space, including spinning binaries and more complex noise conditions. 
Secondly, increasing the expressivity of the generative model and the embedding network should improve accuracy in the most challenging intrinsic parameters while preserving low latency. 
For example, we can use mixture flows, importance sampling corrections, or multiple-stage schemes in which NSF posteriors are refined by targeted \ac{PTMCMC} runs.
Finally, we note that source confusion has not been treated in the present study. Since the embedding network effectively learns a low-dimensional summary of the conditioning data, overlapping GW signals would in general modify the learned summary representation and, correspondingly, the posterior model learned by the normalizing flow. This limitation is not fundamental, however, and the framework can be extended straightforwardly by training on data sets that explicitly include overlapping sources. An alternative strategy would be to first train a source-separation decoder and then retain the current single-event rapid-inference pipeline as a downstream module~\cite{Houba:2025dnr}.

Beyond methodological improvements, the pipeline is also naturally extensible in application, particularly because its architecture is well suited to multiscale signal structure. Although our present demonstration is focused on TianQin, the same framework could be adapted to other space-based detectors, especially LISA~\cite{LISA:2024hlh} and Taiji~\cite{Luo:2021qji}, and potentially also to third-generation ground-based observatories such as the Einstein Telescope~\cite{Branchesi:2023mws} and Cosmic Explorer~\cite{Evans:2021gyd}. A further important step will be to interface 
the GW posteriors with realistic \ac{EM} scheduling and tiling algorithms will be essential for constructing a practical multi-messenger pipeline, in which real-time MBHB inference informs the choice of facilities, filters, and cadences used to search for flare-like \ac{EM} signals. 
Taken together, these developments would move us closer to routine, joint detections of massive black-hole mergers and their \ac{EM} signatures.

\begin{acknowledgments}

X.Z. thanks Jian-dong Zhang and En-Kun Li for helpful discussions on scalability and simulation.
 X.Z. acknowledges support from the Alexander von Humboldt Foundation. Y.H. is supported by the National Key Research and Development Program of China (No. 2023YFC2206700).
 X.Z. also acknowledges support from the Munich Institute for Astro-, Particle and BioPhysics (MIAPbP), funded by the Deutsche Forschungsgemeinschaft (DFG, German Research Foundation) under Germany’s Excellence Strategy (EXC-2094—390783311).

\end{acknowledgments}

\appendix

\section{posterior summaries}

Let $\{\theta_n\}_{n=1}^N$ denote the posterior samples for a given event. Before constructing one-dimensional intervals or identifying sky modes, we apply a robust outlier-suppression step. For one-dimensional parameters, extreme isolated samples are removed using a median-absolute-deviation  filter. For sky-position samples, we construct a local-density proxy from the $k$th-nearest-neighbor distance and remove the low-density tail. This step reduces the impact of a small number of anomalous samples on density estimation and clustering.

For each one-dimensional parameter $x$, we report both an equal-tailed percentile interval and an HPD region. The percentile interval is
\[
[x_{16},x_{84}],
\]
where $x_{16}$ and $x_{84}$ are the $15.865\%$ and $84.135\%$ quantiles. At credibility level $\alpha$, the HPD region is defined by
\[
\int_{\hat p(x)\ge \tau_\alpha}\hat p(x)\,dx=\alpha,
\]
where $\hat p(x)$ is the estimated marginal posterior density. For approximately unimodal marginals, we use a sorting-based shortest-interval estimator; for potentially multimodal marginals, we use a \ac{KDE}- or \ac{KNN}-based density estimate and allow multiple disconnected \ac{HPD} components. We report the total \ac{HPD} width, the number of \ac{HPD} components, and a representative central value given by the density mode or, for the sorting-based estimator, the median.
The \ac{HPD} construction more faithfully captures asymmetric or multimodal posterior structure.

\section{Identification and folding of multimodal sky posteriors}
\label{appendix:c}

Sky-location posteriors in gravitational-wave parameter estimation are significantly multimodal. This multimodality may arise from parameter degeneracies, but can also be enhanced by discrete symmetries of the detector response. As a result, sky-mode identification, mode-center estimation, and credible sky-area evaluation must be performed with care: treating the full posterior as a single connected structure can mix distinct modes and bias the inferred localization uncertainty. In this appendix, we therefore adopt two complementary descriptions. We first introduce a detector-agnostic procedure to identify distinct sky modes and estimate the total sky area by summing the credible areas of all identified modes. We then present a symmetry-based folding construction for TianQin, which provides an alternative and more compact characterization of sky localization after removing discrete symmetry-related degeneracies.

We first identify distinct sky modes in a way that is applicable to generic detectors and generic multimodal sky posteriors. The sky samples are parameterized by longitude $\lambda$ and latitude $\beta$. To avoid the artificial discontinuity at the periodic longitude boundary, we map each sample to the unit sphere,
\[
\mathbf{u}(\lambda,\beta)=
\bigl(
\cos\beta\cos\lambda,\;
\cos\beta\sin\lambda,\;
\sin\beta
\bigr),
\]
and perform clustering in this three-dimensional embedding. After suppressing outliers, we apply DBSCAN and define each non-noise cluster as one sky mode. The total number of modes, $N_{\rm mode}$, is then given by the number of non-noise clusters.

For the $m$th mode $\mathcal{C}_m$, the mode center is defined by the normalized mean unit vector,
\[
\bar{\mathbf{u}}_m=
\frac{\sum_{i\in\mathcal{C}_m}\mathbf{u}_i}
{\left\|\sum_{i\in\mathcal{C}_m}\mathbf{u}_i\right\|},
\]
which is converted back to angular coordinates as
\[
\lambda_m^\star=\operatorname{atan2}(\bar u_y,\bar u_x),\qquad
\beta_m^\star=\arcsin(\bar u_z).
\]
The sky area of this mode at credibility level $\alpha_{\rm sky}$ is defined as the area of its HPD region,
\[
A_m=\int_{\mathcal{R}_m} d\Omega,
\qquad
\int_{\mathcal{R}_m} p_m(\Omega)\,d\Omega=\alpha_{\rm sky},
\]
where $p_m(\Omega)$ is the posterior density renormalized within mode $m$. In practice, we evaluate the area on a HEALPix grid,
\[
\hat A_m = K_\alpha\,\Delta\Omega_{\rm pix},
\qquad
\Delta\Omega_{\rm pix}=\frac{4\pi}{12N_{\rm side}^2},
\]
where $K_\alpha$ is the minimum number of highest-probability pixels satisfying
\[
\sum_{k=1}^{K_\alpha}\hat P_{(k)} \ge \alpha_{\rm sky}.
\]
The total sky area is then obtained by summing over all identified modes,
\[
\hat A_{\rm tot}=\sum_{m=1}^{N_{\rm mode}} \hat A_m.
\]
Throughout this work, we quote sky areas $A_{\rm 90}$ at $\alpha_{\rm sky}=0.9$.

We next introduce a separate characterization tailored to TianQin, based on the discrete symmetries of its sky response. This construction is independent of the multimode clustering-and-summation procedure above and should be regarded as an alternative measure of localization rather than a replacement for it. Because of the detector geometry, a single physical sky location can correspond to multiple symmetry-equivalent solutions in the detector frame, mainly related by reflections with respect to the detector plane and antipodal transformations. Consequently, several posterior sky modes may represent symmetry-related images of the same underlying solution rather than distinct physical modes.

Motivated by this property, we fold symmetry-related samples into a single representative mode. For each original sample $\Omega_i$, we construct its eight symmetry-related images $\{\Omega_i^{(k)}\}_{k=1}^{8}$ according to the discrete symmetry transformations of the TianQin response. We then choose the center of the sample distribution as an anchor point, denoted by $\Omega_{\rm ref}$, and compute the spherical distance from each image to the anchor point,
\begin{equation}
d_i^{(k)} = d_{\rm sph}\!\left(\Omega_i^{(k)},\,\Omega_{\rm ref}\right),
\qquad k=1,\dots,8.
\end{equation}
For each sample, the image with the smallest distance is selected as the folded representative,
\begin{equation}
k_i^\star = \arg\min_{k\in\{1,\dots,8\}} d_i^{(k)},
\qquad
\widetilde{\Omega}_i = \Omega_i^{(k_i^\star)}.
\end{equation}
This defines the folded sample set $\widetilde{\mathcal{S}}=\{\widetilde{\Omega}_i\}$, from which a folded sky area is computed in the same way as for a single mode, denoted by $\hat A_{\rm m,folded}$. If the eight modes are treated as symmetry-equivalent images with comparable areas, the corresponding total folded area may be approximated as
\[
\hat A_{\rm tot,folded}\approx 8\,\hat A_{\rm m,folded}.
\]
Unlike $\hat A_{\rm tot}$, which measures the full multimode localization region based on the clustering algorithm, the folded area provides a compact characterization of the intrinsic angular resolving power after removing the discrete multimode degeneracy induced by the TianQin response.

\section{More testing cases}\label{appendix:a}





To further assess the robustness of the NSF-based pipeline under different mass configurations, and motivated by the fact that the mass-related parameters show somewhat weaker agreement with \ac{PTMCMC} than the sky-localization parameters, we consider two additional representative injections in which only the mass parameters are modified. We denote the reference injection used in Sec.~\ref{sec:acc_latency} as case~I. Keeping all other parameters fixed, we construct two additional cases: in case~II, the chirp mass is changed to $3\times10^5$, yielding an SNR of 178.0; in case~III, the symmetric mass ratio is set to 0.1, yielding an SNR of 123.7.

\begin{figure*}[htbp]
\centering
\includegraphics[width=0.9\textwidth]
{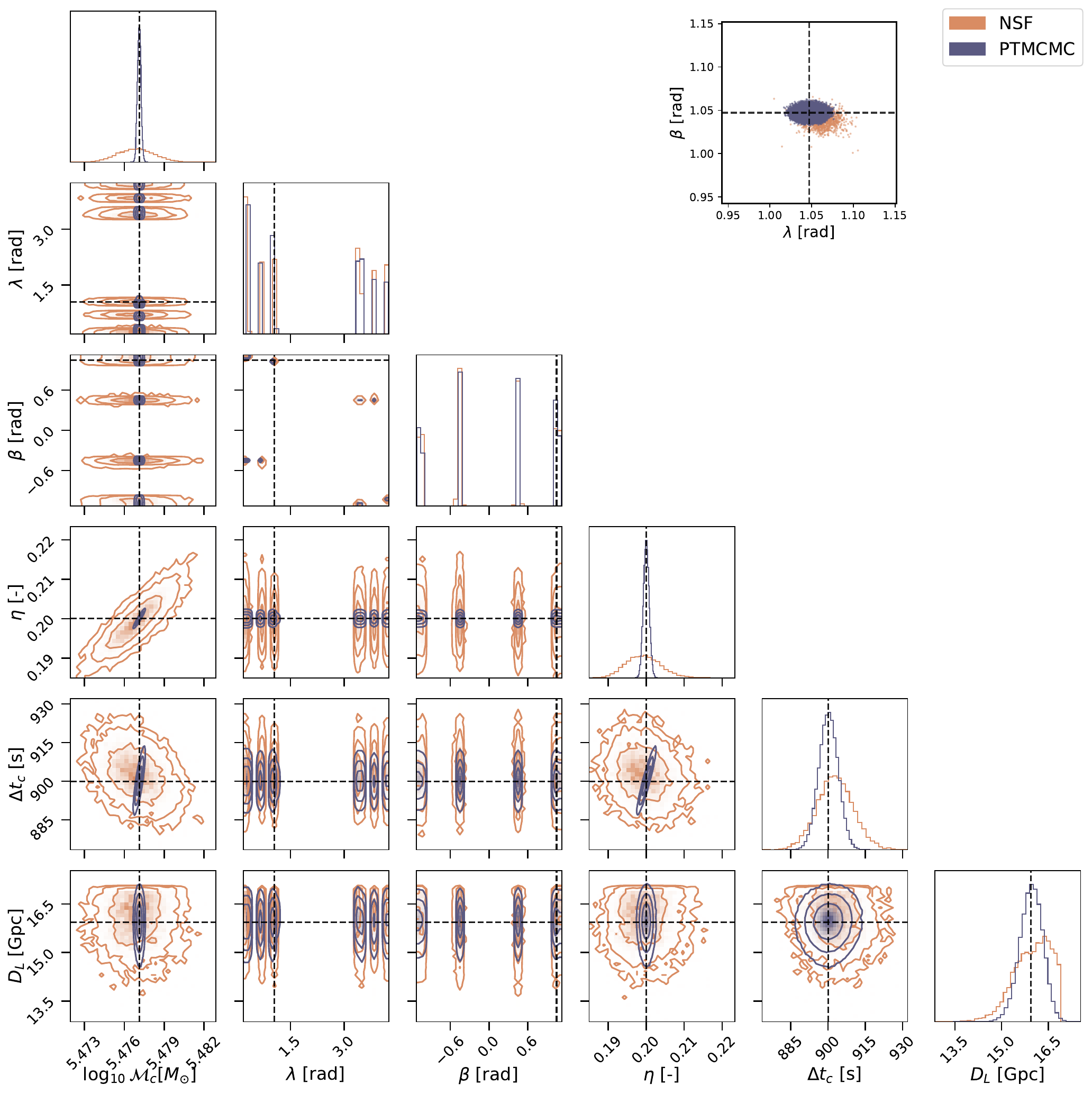}
\caption{Posterior comparison for case~II, in which the chirp mass is modified relative to the reference injection. 
The posteriors obtained with \ac{PTMCMC} (slate blue) and the trained NSF (terracotta) are shown for comparison. 
The dark cross-hairs mark the injected parameter values.
As in the main analysis, \ac{PTMCMC} is evaluated on the noise-free dataset, whereas the NSF is trained and evaluated on data that include detector noise.}
\label{fig:all-params-prediction-case372}
\end{figure*}

\begin{figure*}[htbp]
\centering
\includegraphics[width=0.9\textwidth]
{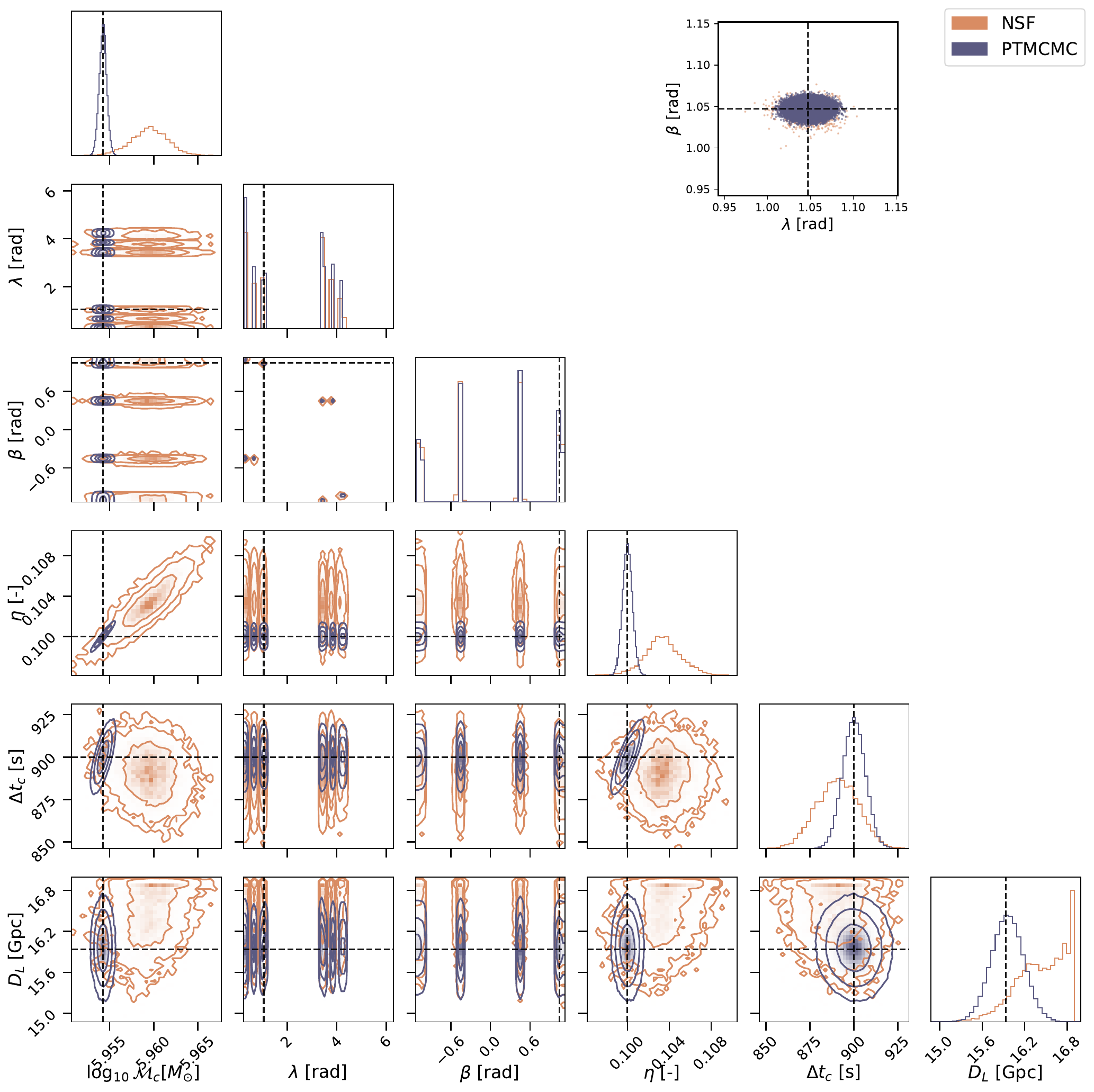}
\caption{Posterior comparison for case~III, in which the symmetric mass ratio is modified relative to the reference injection. 
The posteriors obtained with \ac{PTMCMC} (slate blue) and the trained NSF (terracotta) are shown for comparison. 
The dark cross-hairs mark the injected parameter values. As in the main analysis, \ac{PTMCMC} is evaluated on the noise-free dataset, whereas the NSF is trained and evaluated on data that include detector noise.}
\label{fig:all-params-prediction-case572}
\end{figure*}

Figures~\ref{fig:all-params-prediction-case372} and~\ref{fig:all-params-prediction-case572} show that, for these two additional injections, the NSF and \ac{PTMCMC} recover broadly consistent posterior structures. In particular, for the sky-localization parameters, the NSF continues to yield constraints of a similar order to those obtained in case~I, indicating that the pipeline is comparatively stable in recovering the spatial information most directly relevant for early-warning applications.

A more quantitative comparison is given in Table~\ref{tab:mbhb_pe_both2}. Across all three cases, NSF systematically yields broader posteriors than \ac{PTMCMC} for the mass-related parameters. For example, in case~II the NSF width in $\log M_c$ is $0.0027$, compared with $0.0003$ for \ac{PTMCMC}, while in case~III the corresponding values are $0.0041$ and $0.0008$. For the symmetric mass ratio, the NSF and \ac{PTMCMC} widths are $0.0089$ and $0.0015$ in case~II, and $0.0033$ and $0.0009$ in case~III. This confirms that the NSF posteriors remain noticeably more conservative in the mass-sensitive directions. By comparison, the arrival time and luminosity distance are also generally broader in NSF, but the discrepancy is less pronounced than for the mass parameters, and the resulting constraints remain astrophysically useful for early warning.
At the same time, Table~\ref{tab:mbhb_pe_both2} shows that, for both NSF and \ac{PTMCMC}, all reported fractional uncertainties remain below $10\%$ in all three cases. This is important because it shows that, despite the broader NSF posteriors in several directions, the inferred constraints remain quantitatively useful for astrophysical interpretation and early-warning applications, rather than merely reproducing the posterior structure at a qualitative level.

The corresponding sky-localization statistics are listed in Table~\ref{tab:mbhb_sky_stats2}. In both case~II and case~III, NSF and \ac{PTMCMC} identify $8$ sky modes, showing that the main multimodal structure is recovered robustly. At the same time, the folded angular widths remain of similar order between the two samplers. For instance, in case~II NSF gives $(\Delta\lambda_{\rm m},\Delta\beta_{\rm m})=(0.0270,\,0.0275)$, compared with $(0.0261,\,0.0127)$ for \ac{PTMCMC}$,$ while in case~III the corresponding values are $(0.0559,\,0.0345)$ and $(0.0334,\,0.0168)$. However, NSF generally yields larger clustered and folded sky areas: in case~II, $\hat A_{\rm tot,clustered}=19.57~\mathrm{deg}^2$ for NSF, compared with $4.60~\mathrm{deg}^2$ for \ac{PTMCMC}$,$ and in case~III the corresponding values are $24.09~\mathrm{deg}^2$ and $9.22~\mathrm{deg}^2$. Thus, while NSF recovers the main sky-posterior structure robustly, its sky-area estimates remain systematically more conservative.


Overall, these supplementary experiments show that the NSF-based pipeline can stably recover the sky-localization information most directly relevant for early warning under different mass configurations, while preserving the main posterior structure seen in \ac{PTMCMC}. However, for the mass-related parameters, the NSF posteriors remain significantly broader than those of \ac{PTMCMC}. This indicates that the current method is already practically useful for rapid and robust low-latency localization, while the learning of the posterior in the mass-related directions still leaves room for further improvement.

\begin{table*}[!ht]\caption{Comparison of parameter estimation precision between the NSF and \ac{PTMCMC} in three cases.}
\centering
\begin{tabular}{cccccccccc}
\toprule
\multirow{3}{*}{case} &
\multirow{3}{*}{Sampler} 
 & \multicolumn{7}{c}{Measurement (1$\sigma$ credible interval)}  \\
\cmidrule(lr){3-10}
& 
 & $\Delta \log M_c$   
 & $\frac{\Delta (\log M_c)}{ \log M_c}$ 
 & $\Delta \eta$  
  & $\frac{\Delta \eta}{ \eta}$ 
 & $\Delta t$ 
 & $\frac{\Delta t}{\Delta t_c}$
 & $\Delta D_L$
 & $\frac{\Delta D_L}{D_L}$
 \\
 &
 & [$M_{\odot}$]  
 & [\%] 
 & [-] 
 & [\%]
 &  [Sec] 
  & [\%] 
 &  [Gpc]
 & [\%]
 \\
\midrule
\multirow{2}{*}{case I} &
NSF     
& 0.0037 & 0.0616
& 0.0082 & 4.0881 
& 20.1875  &2.2431
& 1.2281 & 7.7070
\\
& \ac{PTMCMC}  
& 0.0006 & 0.0102 
& 0.0020 & 1.0183
& 13.6224 &1.5136
& 0.6167 &3.8704
\\
\multirow{2}{*}{case II} &
NSF     
& 0.0027 & 0.0500 
& 0.0089 & 4.4730   
& 14.3438  &1.5938
& 1.1693 & 7.3386
\\
& \ac{PTMCMC}  
& 0.0003 & 0.0049 
& 0.0015 & 0.7455
& 7.7277 &0.8586
& 0.7367 &4.6234
\\
\multirow{2}{*}{case III} &
NSF     
& 0.0041 & 0.0685
& 0.0033 & 3.3336 
& 24.031 & 2.6701
& 0.6288 & 3.9467
\\
& \ac{PTMCMC}  
& 0.0008 & 0.0138 
& 0.0009 & 0.9198
& 12.6174 &1.4019
& 0.4538 &2.8480
\\
\bottomrule
\end{tabular}
\label{tab:mbhb_pe_both2}
\end{table*}

\begin{table}[!ht]
\caption{Comparison of sky-localization statistics between the NSF and \ac{PTMCMC} samplers for the same event. Here $N_{\rm mode}$ is the number of sky modes identified by clustering, $\hat A_{\rm tot,clustered}$ is the total sky area obtained by summing over all clustered modes, and $(\Delta\lambda_{\rm m}, \Delta\beta_{\rm m}, \hat A_{\rm m})$ characterize the folded representative mode after removing the discrete symmetry-related degeneracy of the TianQin response. All sky-localization quantities in this table are quoted at the $90\%$ credibility level.}
\centering
\begin{tabular}{clccccc}
\toprule
&
Sampler
& $N_{\rm mode}$
& $\hat A_{\rm tot,clustered}$
& $\Delta\lambda_{\rm m}$
& $\Delta\beta_{\rm m}$
& $\hat A_{\rm m}$ \\
&
& [$-$]
& [deg$^2$]
& [rad]
& [rad]
& [deg$^2$] \\
\midrule
\multirow{2}{*}{case II} &
NSF
& 8
& 19.5660
& 0.0270 & 0.0275
& 2.9375\\
& \ac{PTMCMC}
& 8
& 4.6030
& 0.0261 & 0.0127
& 0.7475 \\
\multirow{2}{*}{case III} &
NSF
& 8
& 24.0903
& 0.0559 & 0.0345
& 4.3538\\
& \ac{PTMCMC}
& 8
& 9.2191
& 0.0334 & 0.0168
& 1.2721\\
\bottomrule
\end{tabular}
\label{tab:mbhb_sky_stats2}
\end{table}


\bibliography{apssamp}

\end{document}